\documentclass[aps,prx,reprint,superscriptaddress,nofootinbib, longbibliography]{revtex4-1}

\usepackage{amsmath,amssymb, amsthm,amstext}
\usepackage{natbib}
\usepackage{graphicx}
\usepackage{color}
\usepackage{array, enumerate}
\usepackage{bm}
\usepackage{multirow}
\usepackage[breaklinks,colorlinks,citecolor=blue]{hyperref}
\usepackage{braket}
\usepackage{txfonts}
\usepackage[normalem]{ulem}
\usepackage{mathtools}

\def\be{\begin{equation}}
\def\ee{\end{equation}}

\def\apjl{Astroph.J.Lett.}
\def\apjs{Astroph.J.S.}
\def\mnras{MNRAS}
\def\araa{ARA\&A}
\def\aap{Astronomy\&Astrophysics}

\begin{document}

\title{Probing orbits of stellar mass objects deep in galactic nuclei with quasi-periodic eruptions -- II: population analysis}
\author{Cong Zhou}
\email{dysania@mail.ustc.edu.cn}
\affiliation{CAS Key Laboratory for Research in Galaxies and Cosmology, Department of Astronomy, University of Science and Technology of China, Hefei 230026, P. R. China}
\affiliation{School of Astronomy and Space Sciences, University of Science and Technology of China, Hefei 230026, P. R. China}
\author{Binyu Zhong}
\affiliation{School of Physics and Technology, Wuhan University, Wuhan 430072, P.R.China}
\author{Yuhe Zeng}
\affiliation{School of Physics, Nankai University, Tianjin 300071, People's Republic of China}
\author{Lei Huang}
\affiliation{Shanghai Astronomical Observatory, Chinese Academy of Sciences, 80 Nandan Road, Shanghai 200030, People’s Republic of China}
\affiliation{Key Laboratory for Research in Galaxies and Cosmology, Chinese Academy of Sciences, Shanghai 200030, People’s Republic of China}
\author{Zhen Pan}
\email{zhpan@sjtu.edu.cn}
\affiliation{Tsung-Dao Lee Institute, Shanghai Jiao-Tong University, Shanghai, 520 Shengrong Road, 201210, People’s Republic of China}
\affiliation{School of Physics \& Astronomy, Shanghai Jiao-Tong University, Shanghai, 800 Dongchuan Road, 200240, People’s Republic of China}

\begin{abstract}
    Quasi-periodic eruptions (QPEs) are intense repeating soft X-ray bursts with recurrence times about a few hours
to a few weeks from  galactic nuclei. Though the debates on the  origin of QPEs have not completely settled down, 
more and more analyses favor the interpretation that QPEs are the result of collisions 
between a stellar mass object (a stellar mass black hole or a main sequence star) and  
an accretion disk around a supermassive black hole (SMBH) in galactic nuclei. 
If this interpretation is correct,
QPEs will be invaluable in probing the orbits of stellar mass objects in the vicinity of SMBHs, 
and further inferring the formation of extreme mass ratio inspirals (EMRIs), one of the major targets of spaceborne gravitational wave missions.
In this work, we extended the EMRI orbital analysis in Paper I \cite{Zhou2024} to all the known QPE sources
with more than $6$ flares observed. Among all the analyzed 5 QPE sources,  two distinct EMRI populations are identified:
4 EMRIs are of low orbital eccentricity (consistent with 0)   which should be born in the wet EMRI formation channel,
and 1 mildly eccentric EMRI  (with $e= 0.25^{+0.18}_{-0.20}$ at 2-$\sigma$ confidence level) is consistent with the predictions of both 
 the dry loss-cone formation channel and the Hills mechanism.

\end{abstract}
\date{\today}

\maketitle

\section{Introduction}
Since the first detection of intense X-ray eruptions more than a decade ago \cite{Sun2013},
X-ray quasi-periodic eruptions (QPEs)  from nine  different nearby galactic nuclei 
have been detected \cite{Miniutti2019,Giustini2020,Arcodia2021,Arcodia2022,Chakraborty2021,Evans2023,Guolo2024,Arcodia2024}.
Most of these QPEs are with recurrence times about a few
hours to ten hours, while QPEs from Swift J023017.0+283603 (Swift J023017 hereafter) 
are exceptional with a much longer recurrence time $\sim$ 3 weeks  \cite{Evans2023,Guolo2024}.
Most of the QPE hosts are dwarf galaxies which harbor low-mass 
($\simeq 10^5-$ a few times $10^6 M_\odot$) central supermassive black holes (SMBHs) \cite{Shu:2017zjd,Wevers2022,Miniutti2023}.
The central SMBH of Swift J023017 is found to be possibly heavier with mass $\log (M_\bullet/M_\odot) = 6.6\pm 0.4$
\cite{Guolo2024}.
The recent detection of QPEs (eRO-QPE 4) from  galaxy 2MASS 04453380-1012047 with a recurrence time $\sim (11 - 15.5)$ hours 
pushed the high-mass end to  $M_\bullet\sim 10^7 M_\odot$  \cite{Arcodia2024}.  Similar to tidal disruption events (TDEs), 
QPEs  are preferentially found in poststarburst galaxies \cite{Wevers2022},
and  three QPE sources (GSN 069,  XMMSL1 J024916.6-04124, eRO-QPE3) and a candidate (AT 2019vcb) have been directly associated with
previous TDEs \cite{Shu2018,Sheng2021,Chakraborty2021,Miniutti2023,Quintin2023,Arcodia2024}.
The presence of a narrow region and the absence of 
luminous broad emission lines in all QPE host galaxies imply the hosts are recently switched-off active galactic nuclei (AGNs) \cite{Wevers2022}.
All the known QPEs are similar in the peak luminosity ($10^{42}-10^{43}$ ergs s$^{-1}$),
the thermal-like X-ray spectra with temperature $kT \simeq 100-250$ eV, and 
the temperature  $50-80$ eV in the quiescent state.
For most QPE sources with more than three flares detected, 
the QPEs display  two alternating peak luminoisities $I_{\rm strong}$ and $I_{\rm weak}$
and two alternating occurrence times $T_{\rm long}$ and $T_{\rm short}$. 
Besides the common properties briefly summarized above and shared by most QPEs, 
there are a number of peculiar features in several QPE sources yielded by long term observations,
including the disappearance and reappearance of QPEs and their association with the quiescent state luminosity, the large change in the  QPE recurrence times $T_{\rm long, short}$,
the complex rising and decay profiles of QPE light curves (see \cite{Miniutti2023b} for GSN 069
and \cite{Arcodia2022,Pasham2024,Chakraborty2024} for eRO-QPE1).

For understanding the origin of QPEs, many models have been proposed, 
including models based on different disk instabilities \cite{Raj2021,PanX2022,Panx2023,Kaur2023,Snieg2023}, 
self-lensing binary massive black hole \cite{Ingram2021},
mass transfer at pericenter from stars or white dwarfs orbiting around the SMBH
in the form of repeating partial tidal disruption events (TDEs) or repeating roche lobe overflows 
\cite{King2020,King2022,King2023,Chen2022,Wang2022,Zhao2022,Metzger2022,Lu2022,Krolik2022,Linial2023b},
and the EMRI+TDE disk model where QPEs are the result of periodic impacts between a stellar-mass object (SMO), 
a main sequence star or a stellar mass black hole (sBH),\footnote{In this model, another  remnant, a neutron star or a white dwarf,
does not work, because its mass is too low compared with a sBH and its radius is too small compared with a main sequence star. Therefore neither the Bondi radius or the geometric radius is sufficient to account for such energetic eruptions (see the energy budget estimates in \cite{Zhou2024} for details).}  
and the accretion disk that is formed following
a recent TDE \cite{Sukova2021,Xian2021,Tagawa2023, Linial2023, Franchini2023}.
As we will see in Section~\ref{subsec:swift}, the accretion disk in some QPE sources is simply a (low-state) AGN accretion disk and seems unrelated to a previous TDE.
Therefore, we will use the more inclusive name “EMRI+disk model'' in this paper.

As briefly summarized above, QPEs are actually bi-quasi-periodic phenomena 
with two different and varying periods $T_{\rm long}$ and $T_{\rm short}$. 
As noted in Ref.~\cite{Zhou2024} (hereafter Paper I), the two varying periods 
are tightly correlated with a stable sum $T_{\rm long}+T_{\rm short}$ for GSN 069.
These two observational facts are the natural consequences in the EMRI+disk model, while pose a huge challenge for other models.
Though the debates on the origin(s) of QPEs have not settled down,  the EMRI+disk interpretation is favored by 
more and more analyses (see e.g., Refs.~\cite{Linial2023,Franchini2023,Arcodia2024,Guolo2024,Zhou2024,Pasham2024,Chakraborty2024,Linial:2024mdz} for detailed discussions).

If the QPEs are indeed sourced by collisions between SMOs and accretion disks around SMBHs, 
they will be a sensitive probe to orbits of SMOs in the vicinity of SMBHs 
and consequently to the formation rate and formation processes of EMRIs \cite{Zhou2024,Arcodia:2024efe,Kejriwal:2024bna}. 
As shown in Paper I, the EMRI orbital parameters can be constrained from the flare timing of QPEs,
and the EMRI in GSN 069 is found to be of low eccentricity ($e \lesssim 0.1$)
and semi-major axis about $\mathcal{O}(10^2)$ gravitational radii of the central SMBH.
This nearly circular and tight EMRI orbit provides a strong constraint on its formation channel:
it is consistent with the prediction of the (wet) AGN  disk channel \cite{Sigl2007,Levin2007,Pan2021prd,Pan2021b,Pan2021,Pan2022,Derdzinski2023,Wang2023,Wang2023b}, 
while incompatible with
the (dry) loss-cone channel \cite{Hopman2005,Preto2010,Bar-Or2016,Babak2017,Amaro2018,Broggi2022} or the Hills mechanism \cite{Miller2005,Raveh2021}.
In this work, we apply the same orbital analysis to all the known QPEs with more than 6 flares detected where the EMRI orbit
can be reasonably constrained.
Among all the 5 analyzed QPE sources, we find that 4 EMRIs are of 
low eccentricity (consistent with 0, see Figs.~\ref{fig:GSN_069_corner}-\ref{fig:Swift_corner}) 
which are consistent with the prediction of the wet EMRI formation channel,
and 1 EMRI with non-zero eccentricity ($e= 0.25^{+0.18}_{-0.20}$ at 2-$\sigma$ confidence level) is consistent with the predictions of both the Hills mechanism and 
the dry loss-cone formation channel. If these EMRIs are a representative sample of sBH EMRIs,
we expect to see a  bimodal distribution of EMRI eccentricities 
as they enter the sensitivity band of spaceborne gravitational detectors (see Fig.~\ref{fig:e-p_statistics}).

This paper is organized as follows. In Section~\ref{sec:model}, we briefly review the EMRI+disk model, 
and the main points of the Bayesian analysis method we will use for constraining the model parameters.
In Section~\ref{sec:obt_par}, we show the EMRI orbital parameters inferred from the QPE light curves.
In Section~\ref{sec:summary}, we discuss the implications of QPEs on the EMRI formation and future gravitational wave (GW) detection of EMRIs.
In Appendix, we summarize the posterior corner plot of the flare timing model parameters of each QPE source analyzed in this work.
Throughout this paper, we use the geometrical units with convention $G=c=1$.

\section{EMRI+disk model}\label{sec:model}

As discussed in Paper I,  given an EMRI+disk system, one can in principle predict
both the collision times between the SMO and the disk and the resulting QPE light curve.  
For mitigating the impact of the uncertainties in the disk model, we choose to 
constrain the EMRI kinematics and the QPE emission separately: we first fit each QPE with a simple light curve model
and obtain the starting time of each flare $t_0 \pm \sigma(t_0)$, which is
identified as the observed disk crossing time and used for constraining the
EMRI kinematics.

According to the Bayes theorem, the posterior of parameters given data is
\be \mathcal P(\mathbf{\Theta}|d) \propto \mathcal{L}(d|\mathbf{\Theta}) \pi(\mathbf{\Theta})\ , \ee
where $\mathcal{L}(d|\mathbf{\Theta})$ is the likelihood of detecting data $d$ given a model with parameters $\mathbf{\Theta}$ and $\pi(\mathbf{\Theta})$ is the parameter prior assumed.
In the following two subsections, we will explain the light curve model, the flare timing model, and define their likelihoods.

\subsection{Light curve model}
In Paper I, we considered two light curve models, a physical plasma ball model and a phenomenological model, in analyzing the GSN 069 flares.
As a result, we obtained consistent constraints on the flare starting times with the two models and therefore consistent constraints of the EMRI orbital parameters.
For simplicity in this work, we only use the simple phenomenological model \cite{Norris2005, Arcodia2022}
\be \label{eq:phen}
L_{\rm X}(t) =
    \begin{cases}
     0  & \text{if $t\leq t_p-t_{\rm as}$}\\
      L_p e^{\sqrt{\tau_1/\tau_2}} e^{\tau_1/(t_p-t_{\rm as}-t)} & \text{if $t_p-t_{\rm as} < t<t_p$}\\
      L_p e^{-(t-t_p)/\tau_2} & \text{if $t\geq t_p$}
    \end{cases}       
\ee 
in fitting the flare light curves, 
where $t_{\rm as}=\sqrt{\tau_1\tau_2}$. Following the definition in Ref.~\cite{Norris2005}, 
we define the flare  starting time as when the flux is $1/e^3$ of the peak value, i.e.,  $L_{\rm X}(t_0) = L_p/e^3$.
Therefore only 3 out of the 4 time variables are independent, and we take $\{t_0, t_p, \tau_2\}$ and $L_p$ 
as the independent model parameters.

In addition to the QPEs, quasi-periodic oscillations (QPOs)
has been identified in the quiescent state luminosity of  GSN 069, we therefore model the background luminosity as
\be  \label{eq:bgd}
L_{\rm bgd} (t)=B+A \sin\left(2\pi (t-t_0)/P_{\rm QPO} + \phi_{\rm QPO}\right)\ ,
\ee 
where $B$ is the average background luminosity, $A, P_{\rm QPO}, \phi_{\rm QPO}$ are the QPO amplitude, period and initial phase, respectively. 
For other QPE sources considered in this work, there is no clear signature of QPOs, therefore the background luminosity is simply modeled 
as a constant $L_{\rm bgd} (t)=B$.

Similar to in Paper I, we define the likelihood of seeing data $d$ given the light curve model above as 
\be 
 \mathcal{L}_{\rm emission}(d|\mathbf{\Theta}) = \prod_{i}\frac{1}{\sqrt{2\pi (F\sigma_i)^2}}
 \exp\left\{-\frac{(L(t_i)-d_i)^2}{2(F\sigma_i)^2} \right\} \ ,
\ee 
where $d_i, \sigma_i$ are the measured luminosity and errorbar at time $t_i$, respectively,  
$L(t_i) = L_X(t_i) + L_{\rm bgd}(t_i)$ is the model predicted luminosity [Eqs.~(\ref{eq:phen},\ref{eq:bgd})], and $F$ is a scale factor taking possible calibration uncertainty into account.
For Swift J023017 in quiescent state, non-detection has been reported in the form of upper limits, e.g., $\mu(t_j)$ at $99\%$ confidence level.
The likelihood of seeing these upper limits can be formulated as 
\be 
 \mathcal{L}_{\rm emission}^{\rm non-det}(d|\mathbf{\Theta}) = \prod_{j} \frac{1}{2F\mu_j} {\rm erfc}\left(\frac{L(t_j)-F\mu_j}{0.036F\mu_j} \right) \ ,
\ee 
where ${\rm erfc}(x)$ is the complementary error function and the factor $0.036$ comes from the 
confidence level constraint $\int_1^\infty \frac{1}{2}{\rm erfc}\left(\frac{x-1}{0.036}\right) dx = 0.01$.
In the similar way, the scale factor $F$ has been introduced taking possible calibration uncertainty into account.

\subsection{Flare timing model}
In the EMRI+ disk model, we assume a SMO moving on the geodesic and a steady disk lying on the equator, 
and the observed flare starting time is identified as when the SMO crosses the upper surface of the accretion disk 
plus light propagation delays to the observer.
In this section, we will first introduce the basics of 
EMRI kinematics around a Kerr black hole (BH), then the propagation delays and finally define the likelihood function.

In the Kerr spacetime, the EMRI geodesic is integrable with four integrals of motion, $\{E, L, C, \mathcal{H}\}$.
In term of 4-momentum components $p_\mu$, the energy and the $z$-component of angular momentum 
are 
\be 
E = -p_t\ , 
\ee 
and 
\be 
L = p_\phi\ .
\ee 
The Carter constant is \cite{Carter1968}
\be \label{eq:Carter}
C = p_\theta^2 + a^2\cos^2\theta(\mu^2-p_t^2)+\cot^2\theta p_\phi^2\ ,
\ee 
and the Hamiltonian is 
\be 
\begin{aligned}
   \mathcal{H} = \frac{1}{2}g^{\mu\nu}p_\mu p_\nu = &\frac{\Delta}{2\Sigma}p_r^2 + \frac{1}{2\Sigma} p_\theta^2
+\frac{(p_\phi+a\sin^2\theta p_t)^2}{2\Sigma\sin^2\theta} \\
&-\frac{[(r^2+a^2)p_t+ap_\phi]^2}{2\Sigma\Delta}\ ,
\end{aligned}
\ee 
where $\mu$ is the rest mass of the SMO, $a$ is the dimensionless spin, $\Delta = r^2-2r+a^2$, and $\Sigma = r^2+a^2\cos^2\theta$.
Inverting the four equations above, the momentum $p_\mu$ can be written in terms of the integrals of motion as 
\be \label{eq:momenta}
\begin{aligned}
    p_r &= \pm\frac{\sqrt{V_r(r)}}{\Delta}\ , &p_\theta &= \pm \sqrt{V_\theta(\theta)} \ , \\ 
    p_t &= -E\ , &p_\phi &= L \ ,
\end{aligned}
\ee 
where the two potentials are  
\be 
\begin{aligned}
    V_r(r) &= [(r^2+a^2)E-aL]^2-\Delta[\mu^2r^2+(L-aE)^2+C]\ , \\
    V_\theta(\theta) &= C-\left[(\mu^2-E^2)a^2+\frac{L^2}{\sin^2\theta}\right]\cos^2\theta\ .
\end{aligned}
\ee 
These two potentials define the boundaries of the geodesic motion in the $r$ and $\theta$ directions, respectively:
the pericenter/apocenter separations $r_{-/+}$
are the two largest roots to equation $V_r(r)=0$, and the minimum polar angle $\theta_{\rm min}$
is the root to equation equation $V_\theta(\theta)=0$.
In terms of commonly used orbital parameters, eccentricity $e$ and semi-latus rectum $p$, $r_\pm=p/(1\mp e)$.
The conversion relations between the integrals of motion $(E, L, C)$ and the  orbital parameters $(p, e, \theta_{\rm min})$ has been derived in Ref.~\cite{Schmidt2002}.

In the Hamiltoninan language, the equations of motion are 
in the familiar form 
\be 
\begin{aligned}
    \dot x^\mu &= \frac{\partial \mathcal{H}}{\partial p_\mu}\ , \\ 
    \dot p_\mu &= - \frac{\partial \mathcal{H}}{\partial x^\mu}\ ,
\end{aligned}
\ee 
where the dot is the derivative with respect to  the proper time.
Considering a bound orbit with parameters $(p, e, \theta_{\rm min})$, we first obtain the integrals of motion $(E, L, C)$ using the conversion relation \cite{Schmidt2002}, then set the initial condition
\be 
\begin{aligned}
    t_{\rm ini} &= t_0^{(1)}\ , \\ 
    r_{\rm ini} &= \frac{p}{1+e\cos\chi_{r0}}\ ,  \\ 
    \cos\theta _{\rm ini} &= \cos\theta_{\rm min} \cos\chi_{\theta0}\ , \\ 
    \phi_{\rm ini} &= \phi_0\ , \\
    p_{r,\rm ini} &= \frac{\sqrt{V_r(r_{\rm ini})}}{\Delta} {\rm sign}(\sin\chi_{r0})\ ,\\ 
    p_{\theta,\rm ini} &= \sqrt{V_\theta(\theta_{\rm ini})}{\rm sign}(\sin\chi_{\theta 0})\ ,
\end{aligned}
\ee 
where $t_0^{(1)}$ is the starting time of the 1st flare observed,
and $\chi_{r0},\chi_{\theta 0}, \phi_0$ are the initial phases in the $r,\theta, \phi$ directions, respectively.

Without loss of generality, we set the 
line of sight angles as $\theta_{\rm obs}\in (0, \pi/2), \phi_{\rm obs}=0$, i.e., the observer
lies in the $x-z$ plane with unit direction vector $\vec n_{\rm obs} = (\sin\theta_{\rm obs}, 0, \cos\theta_{\rm obs})$,
consequently observable collisions happen when the SMO crosses the upper surface of the disk $z(t_{\rm crs})= H$
(we set the disk height to the mid-plane as $H=1.5 M_\bullet$ in this work). 
The propagation times of different flares at different collision locations $r_{\rm crs} \vec n_{\rm crs}$
to the observer will also be different. Taking the light propagation delays into account,
we can write $t_{\rm obs} = t_{\rm crs} + \delta t_{\rm geom} + \delta t_{\rm shap}$, where 
\be\label{eq:tobs}
\begin{aligned}
    \delta t_{\rm geom} &= -r_{\rm crs} \vec n_{\rm obs}\cdot \vec n_{\rm crs}\ , \\ 
    \delta t_{\rm shap} &=-2M_\bullet\ln[r_{\rm crs} (1+ \vec n_{\rm obs}\cdot \vec n_{\rm crs})]\ ,
\end{aligned}
\ee 
are corrections caused by different path lengths and different Shapiro delays \cite{Shapiro1964}, respectively.

With model predicted flare starting times $t_{\rm obs}(\mathbf{\Theta})$ and the data $d=\{t_0^{(k)} \pm \sigma(t_0^{(k)})\}$ obtained from fitting the QPE light curves,
the likelihood of seeing data $d$ in the EMRI+disk model with parameters  $\mathbf{\Theta}$ is therefore
\be 
 \mathcal{L}_{\rm timing}(d|{\mathbf{\Theta}}) = \prod_{k}\frac{1}{\sqrt{2\pi (\sigma(t_0^{(k)}))^2}}
 \exp\left\{\frac{(t_{\rm obs}^{(k)}- t_0^{(k)})^2}{2(\sigma(t_0^{(k)}))^2} \right\} \ .
\ee 
As noticed in Paper I,  the simple phenomenological model [Eq.~(\ref{eq:phen})] 
may not capture all physical processes that contribute to the QPE light curves.
To model the possible extra time delays/advances from these unmodeled processes, we inflated the uncertainties $\sigma(t_0^{(k)})$ by a free factor $F_t$ in Paper I. A more motivated approach of delineating unmodeled effects has been proposed in the context of 
hierarchical test of  General Relativity with gravitational waves \cite{Isi2019}. In the same approach, we assume the unmodeled delays follows a Gaussian distribution 
\be 
p(\delta t_{\rm sys}^{(k)}) = \frac{1}{\sqrt{2\pi \sigma_{\rm sys}^2}}  \exp\left(-\frac{(\delta t_{\rm sys}^{(k)})^2}{2 \sigma_{\rm sys}^2}\right)\ ,
\ee 
and the likelihood is modified as 
\be 
\begin{aligned}
    \mathcal{L}_{\rm timing}(d|{\mathbf{\Theta}}) 
    &= \prod_{k}\int \frac{1}{\sqrt{2\pi (\sigma(t_0^{(k)}))^2}}
 \exp\left\{-\frac{(t_{\rm obs}^{(k)}+\delta t_{\rm sys}^{(k)}- t_0^{(k)})^2}{2(\sigma(t_0^{(k)}))^2} \right\} \\ 
 &\times p(\delta t_{\rm sys}^{(k)}) \ d \delta t_{\rm sys}^{(k)} \ ,\\
 &=\prod_{k}\frac{1}{\sqrt{2\pi (\tilde\sigma(t_0^{(k)}))^2}}
 \exp\left\{-\frac{(t_{\rm obs}^{(k)}- t_0^{(k)})^2}{2(\tilde\sigma(t_0^{(k)}))^2} \right\} \ ,
\end{aligned}
\ee 
where $(\tilde\sigma(t_0^{(k)}))^2=(\sigma(t_0^{(k)}))^2+\sigma_{\rm sys}^2$ is the uncertainty contributed by both modeled and unmodeled uncertainties.

In addition to the QPE light curves, the central SMBH masses are inferred from the stellar velocity dispersion using the $M_\bullet-\sigma_\star$ relation \cite{Tremaine2002,Gultekin2009}.
The likelihood is naturally defined as 
\be 
\mathcal{L}_{M_\bullet}(d|{\mathbf{\Theta}}) = \frac{1}{\sqrt{2\pi \sigma_{\log_{10} M_\bullet}^2}} \exp\left\{-\frac{(\log_{10} M_\bullet - \mu_{\log_{10} M_\bullet})^2}{2 \sigma_{\log_{10} M_\bullet}^2} \right\}\ ,
\ee 
where $\mu_{\log_{10} M_\bullet}$ and $\sigma_{\log_{10} M_\bullet}$ are the central value and the uncertainty of inferred SMBH mass, respectively (see \cite{Wevers2022} for a brief summary of the mass measurements of SMBHs in the QPE host galaxies).

To summarize, 
there are $10$ parameters in the flare timing model with the total likelihood $\mathcal{L}_{\rm timing}\times\mathcal{L}_{M_\bullet}$: 
the intrinsic orbital parameters $(p, e, \theta_{\rm min})$, the initial phases $(\chi_{r0}, \chi_{\theta0}, \phi_0)$,
the polar angle of the observer $\theta_{\rm obs}$, the mass of the SMBH $M_\bullet$ or equivalently the orbital period $T_{\rm obt} :=2\pi (A/M_\bullet)^{3/2} M_\bullet$  (with semi-major axis $A = p/(1-e^2)$), 
the dimensionless spin of the SMBH $a$ and the parameter $\sigma_{\rm sys}$.
Same to Paper I, the model parameter inferences in this work are also performed using the \texttt{nessai} \cite{nessai} algorithm in  \texttt{Bilby} \cite{Ashton2019}.

\section{EMRI orbital parameters}\label{sec:obt_par}
Currently, eight QPE sources and one candidate in total \cite{Miniutti2019,Giustini2020,Arcodia2021,Arcodia2022,Chakraborty2021,Evans2023,Guolo2024,Arcodia2024,Quintin2023}
have been reported. Not all are useful to our EMRI orbital analysis due to the limited number of flares detected from some of these sources. 
For XMMSL1 J024916.6-04124 \cite{Chakraborty2021}, only 1.5 flares 
have been detected, with a flare fully resolved and another partially resolved at the edge of the exposure time. For AT 2019vcb \cite{Quintin2023}, only 1 flare has been reported,
therefore this source is classified as a candidate. For both eRO-QPE 3 and 4 \cite{Arcodia2024}, only 3 flares have been detected by XMM-Newton (while NICER data are of bad quality in solving the flares).

In this work, we apply the EMRI orbital analysis to the remaining 5 QPE sources, where a reasonable number of flares are available for constraining the model parameters.
For GSN 069, we use the same light curve data as in Paper I (see Paper I for more data reprocessing details). For other sources that have been detected with XMM-Newton, we reprocessed the raw data from EPIC-pn camera \cite{EPIC-pn} of the XMM-Newton mission, using the latest XMM–Newton Science Analysis System (SAS) and the Current Calibration Files (CCF). The photon arrival times are all barycentre-corrected in the DE405-ICRS reference system. For eRO-QPE 1, we directly take the data in Ref.~\cite{Chakraborty2024}. For Swift J023017, we take the data in Ref.~\cite{Guolo2024}. In the remainder of this section, we first briefly introduce the observations available of each QPE source, 
then summarize the constraints on the orbital parameters obtained from these observations.

\subsection{GSN 069}\label{subsec:GSN069}

\begin{figure*}
\includegraphics[scale=0.58]{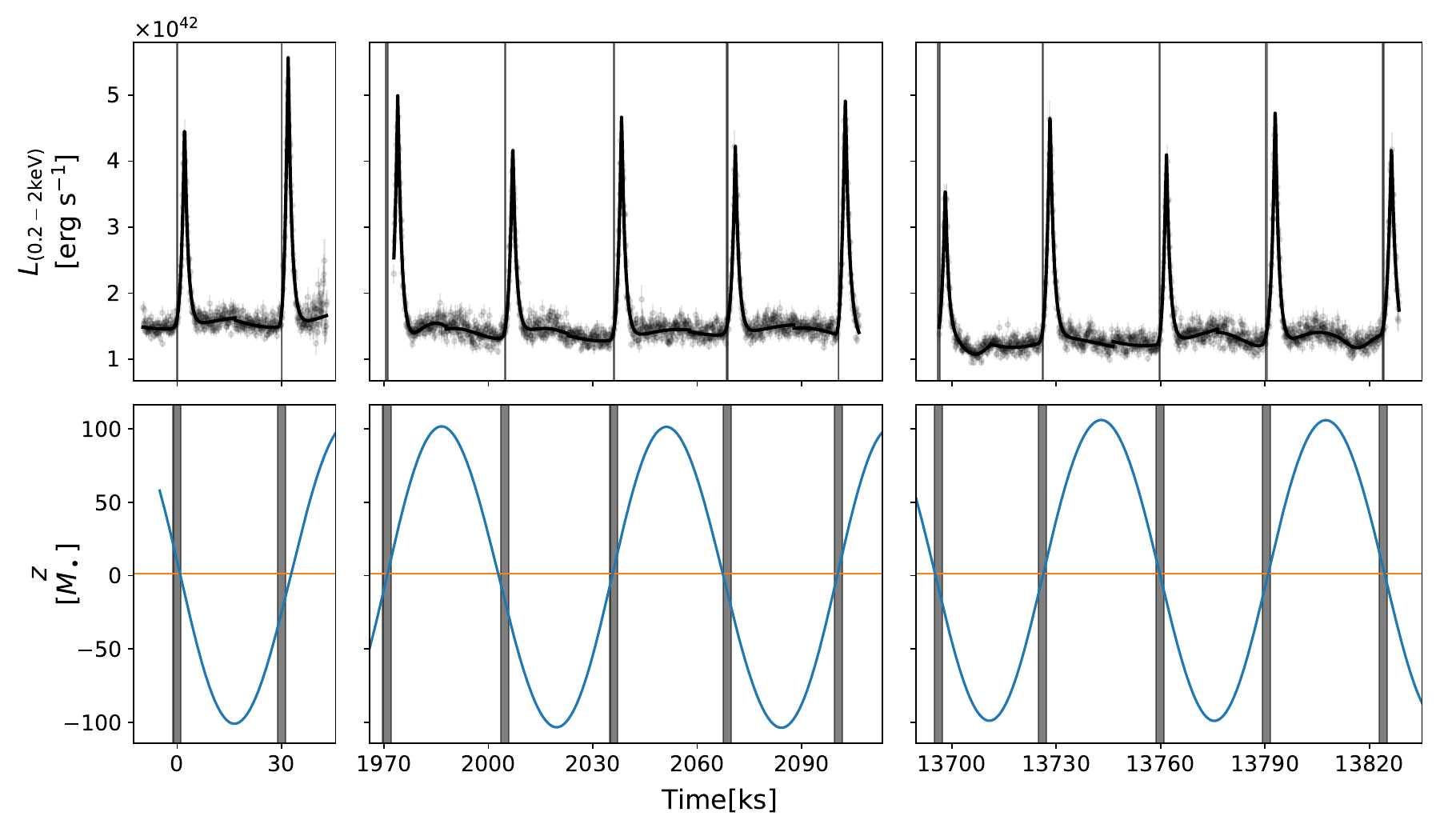}
\caption{\label{fig:GSN_069_lc} Top panel: light curve data of GSN 069 along with the best fit of the light curve model, where the vertical bands are the inferred starting times $t_0^{(k)}\pm \sigma(t_0^{(k)})$ of the QPEs. Bottom panel: $z(t)$ of the best-fit orbit ($p=185 M_\bullet, e=0.04, T_{\rm obt}= 64.6\ {\rm ks}$ ), where the horizontal line denotes the disk surface $z=H$ and the verticals bands are the inferred starting times $t_0^{(k)}\pm \tilde\sigma(t_0^{(k)})$, with
$\tilde\sigma(t_0^{(k)})=\sqrt{(\sigma(t_0^{(k)}))^2+\sigma_{\rm sys}^2}$.}
\end{figure*}

GSN 069 have been extensively monitored (XMM Newton 1-12 and Chandra)  since the first detection of QPEs a decade ago
\cite{Miniutti2023,Miniutti2023b}.
QPEs are found only in XMM 3-6 and 12 and the Chandra observation, when the quiescent state luminosity is low. 
From the quiescent state light curves, it is likely that two (partial) TDEs has happened with $\sim 9$ years apart.
QPEs in XMM 3-5 shows clear alternating long-short pattern in the recurrence times and alternating strong-weak pattern 
in the QPE intensities. During XMM 6 (and possibly 12), the QPEs become irregular in the sense
that the alternating strong-weak pattern is not well preserved. 
This is likely the result of the perturbation from the second partial TDE (see Paper I for more details).
Therefore we use observations XMM 3-5 only for constraining the EMRI orbital parameters as in Paper I.
The central SMBH mass inferred from the stellar velocity dispersion measurement is $\log_{10}(M_\bullet/M_\odot) = 5.99\pm 0.5$ \cite{Saxton2011}.

In Table~\ref{tab:GSN069}, we show the priors used in the Bayesian inference of orbital parameters.
All of them are uninformative uniform priors in a reasonable range.

\begin{table}
    \centering
    \resizebox{0.4\columnwidth}{!}{%
    \begin{tabular}{l|ccc}
       $\mathbf{\Theta}$ & $\pi(\mathbf{\Theta})$ &  &  \\
        \hline
          $p\ [M_\bullet]$ & $\mathcal{U}[50, 500]$ &  &  \\ 
         $e$ &$\mathcal{U}[0, 0.9]$ &  & \\
       $\cos(\theta_{\rm min})$ & $\mathcal{U}[0, 1]$&  & \\
       $\chi_{r,0}$ & $\mathcal{U}[0, 2\pi]$ &  & \\
       $\chi_{\theta,0}$ &$\mathcal{U}[0, \pi]$ &  & \\
       $\phi_{0}$ & $\mathcal{U}[0, 2\pi]$ &  & \\
       $T_{\rm obt}\ [{\rm ks}]$ & $\mathcal{U}[60, 70]$ &  & \\
       $a$ & $\mathcal{U}[0, 1]$&  & \\
       $\theta_{\rm obs}$ & $\mathcal{U}[0, \pi/2]$  &  & \\
       $\sigma_{\rm sys} \ [{\rm ks}]$ & $\mathcal{U}[0, 2]$  &  &
    \end{tabular} }
    \caption{The priors used in the orbital parameter inference of GSN 069 EMRI. The priors of different sources will be slightly adjusted making sure the parameter posteriors are informed by the data.} 
    \label{tab:GSN069}
\end{table}

In Fig.~\ref{fig:GSN_069_lc}, we show the EMRI orbit of the best-fit flare timing model (with orbital parameters $p=185 \ M_\bullet, e=0.04, T_{\rm obt}=64.6 \ {\rm ks}$ )
along with the starting time of each flare. 
The posterior corner plot of all the model parameters is shown in Fig.~\ref{fig:GSN_069_corner} in Appendix, where the orbital parameters are constrained as
\be 
\begin{aligned}
    p &= 279^{+188}_{-182} \ M_\bullet\ , \\ 
    e &= 0.05^{+0.09}_{-0.05} \ ,\\
    T_{\rm obt} &= 64.6^{+1.1}_{-1.9}\ {\rm ks}\ ,
\end{aligned}
\ee 
at 2-$\sigma$ confidence level, respectively. 
In addition, the SMBH spin $a$ is unconstrained because the Lense-Thirring precession timescale of the SMO is too long to be resolved in the limited observations of GSN 069.
Not surprisingly, the constraints above are wider compared to and consistent with those in Paper I, where a Schwarzschild SMBH ($a=0$) is assumed.

In comparison with the best-fit orbital parameters in Paper I ($p = 171 M_\bullet, e=0.04, T_{\rm obt} = 63.7$ ks), there is a $\delta T_{\rm obt}=0.9$ ks difference in the two orbital periods. As a result, the two orbits starting with the same initial phase (XMM 3) are off by half cycle at $t=T_{\rm obt}^2/(2\delta T_{\rm obt})\approx 2\times 10^3$ ks (XMM 4).

Though the alternating long-short pattern in the flare recurrence times is usually attributed to the non-zero orbital eccentricity, 
it is actually also possible to observe such pattern in the case of a circular SMO orbit because of 
different light propagation delays from the two collision locations in one orbital period to the observer (see Eq.~(\ref{eq:tobs})).
That's why the orbital eccentricities of GSN 069 and a few other QPE sources  with 
alternating long-short pattern in the flare recurrence times are found to be consistent with zero.

\subsection{RX J1301.9+2747}

\begin{figure*}
\includegraphics[scale=0.58]{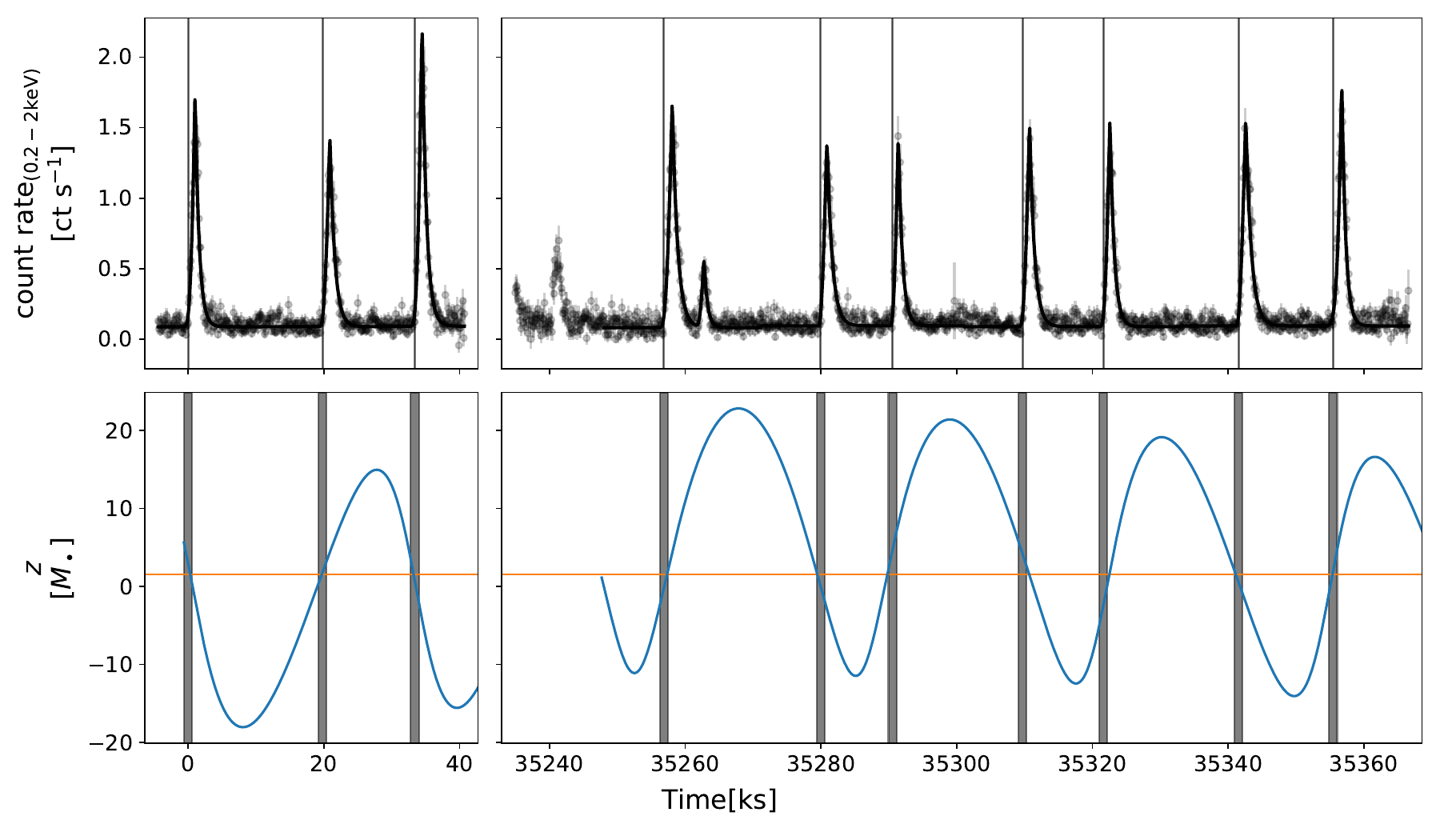}
\caption{\label{fig:RX_lc} Same to Fig.~\ref{fig:GSN_069_lc} except for RX J1301.9+2747 with best-fit orbital parameters ($p=36 M_\bullet, e=0.35, T_{\rm obt}= 31.7\ {\rm ks}$),
where the origin of the tiny flare at $t\approx 35265$ ks is unclear therefore is not included in the EMRI orbital analysis.
From the best-fit orbit, we clearly see the orbital apsidal precession.}
\end{figure*}

RX J1301.9+2747 (hereafter RX J1301) is the second identified but the first detected QPE source. In 1991, a rapid flare, albeit only decaying phase, was detected in the ROSAT light curve of RX J1301.9+2747 \cite{Dewangan2000}. In 2000, the first complete flare detection of RX J1301 was made by XMM-Newton, where a single flare was seen in EPIC-PN and 1.5 flares were seen in EPIC-MOS \cite{EPIC-mos}. A similar single flare was detected by Chandra nine years later \cite{Sun2013, Giustini2020}. 
It is following the discovery of the first QPE source, GSN 069, that RX J1301 was identified as a newly recognized QPE source through a literature scan \cite{Giustini2020}. 

Four recent XMM-Newton observations of RX J1301 are available (XMM 1-4). In XMM 1 (2019-05-30) and 2 (2020-07-11), the QPEs are in a regular phase similar to that of GSN 069, showing alternating long-short pattern in the recurrence times and
alternating strong-weak pattern in the QPE intensities.  In XMM 3 (2022-06-17) and 4 (2022-06-19), the QPEs become irregular with a large variation in the recurrence time within one day.
Similar to in the GSN 069 case, we apply the EMRI orbital analysis only using the regular-phase observations XMM 1-2.

The priors used are the same as the GSN 069 ones (Table~\ref{tab:GSN069}) except $\mathbf{\pi}(p)= \mathcal{U}[5, 500] \ M_\bullet$, $\mathbf{\pi}(T_{\rm obt}) = \mathcal{U}[10, 70]$ ks, $\mathbf{\pi}(\sigma_{\rm sys})=\mathcal{U}[0, 3]$ ks. The central SMBH mass inferred from the stellar velocity dispersion measurement is $\log_{10}(M_\bullet/M_\odot) = 6.65 \pm 0.42$ \cite{Wevers2022}.

In Fig.~\ref{fig:RX_lc}, we show the EMRI orbit of the best-fit flare timing model (with orbital parameters $p=36 \ M_\bullet, e=0.35, T_{\rm obt}=31.7 \ {\rm ks})$
along with the starting time of each flare. 
The posterior corner plot of all the model parameters is shown in Fig.~\ref{fig:RX_corner}, where the orbital parameters are constrained as
\be 
\begin{aligned}
    p &= 52^{+80}_{-30} \ M_\bullet\ , \\ 
    e &= 0.25^{+0.18}_{-0.20} \ , \\
    T_{\rm obt} &= 32.0^{+0.9}_{-1.0}\ {\rm ks}\ ,
\end{aligned}
\ee 
at 2-$\sigma$ confidence level, respectively. The EMRI in this source is confidently non-circular, and is quite different from EMRIs in other QPE sources. The EMRI formation analysis in the following section shows that the RX J1301 EMRI may be born in a different formation channel  from other EMRIs.

The tight orbit itself also constrains the SMO nature mildly. 
If the SMO is a main sequence star, it should be free from (partial) tidal disruption by the central SMBH before the QPEs turn on, 
i.e., the star orbit should satisfy the following stability condition
\be \label{eq:RX_stable}
\frac{r_{\rm p}}{2 r_{\rm t}}\approx  1.07(1-e) \   T_{\rm obt, 32}^{2/3}  R_{\star,\odot}^{-1} m_{\star,\odot}^{1/3} > 1\ ,
\ee 
where $m_{\star,\odot} :=m_\star/m_\odot, R_{\star,\odot} :=R_\star/R_\odot$, $T_{\rm obt, 32}:=T_{\rm obt}/32$ ks,
$r_{\rm p}$ is the pericenter distance and $r_{\rm t} := (M_\bullet/m_\star)^{1/3} R_\star$ is the tidal disruption radius.
Note that $R_\star \propto m_\star^{0.8}$ for main sequence stars.
The above stability condition requires a low eccentricity orbit and/or a sub-solar mass star:
$e < 0.07$ for $m_{\star, \odot} = 1$; $m_{\star, \odot} < 0.46 $ for $e=0.35$ (the best-fit value); 
$m_{\star, \odot} < 0.23 $ for $e=0.43$ (the 2-$\sigma$ boundary).

\subsection{eRO-QPE 1}

\begin{figure*}
\includegraphics[scale=0.58]{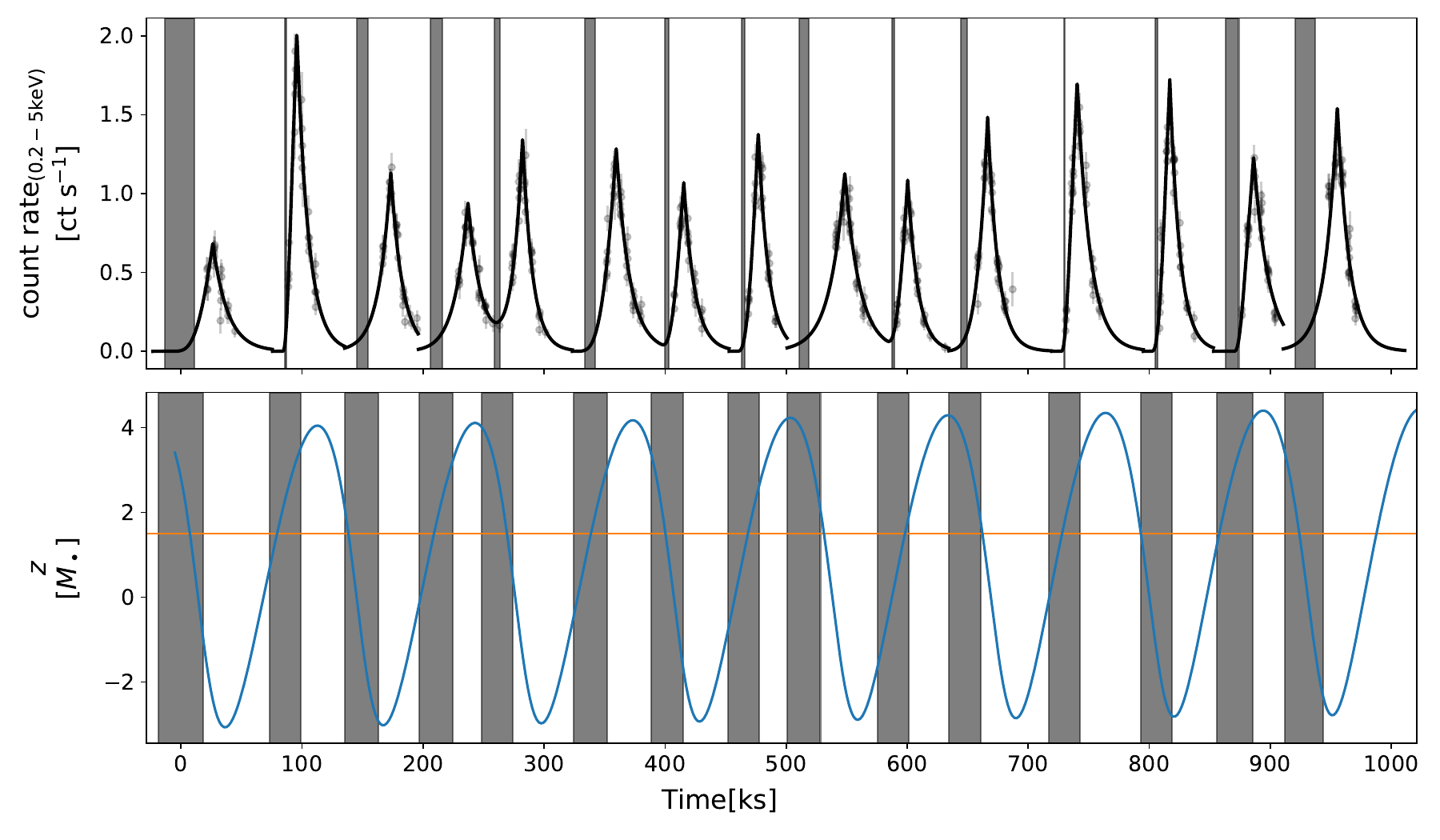}
\caption{\label{fig:eRO1_202008_lc} Same to Fig.~\ref{fig:GSN_069_lc} except for eRO-QPE 1 (2020-08) with best-fit orbital parameters ($p=322 M_\bullet, e=0.30, T_{\rm obt}= 130.6\ {\rm ks}$).}
\end{figure*}

\begin{figure*}
\includegraphics[scale=0.58]{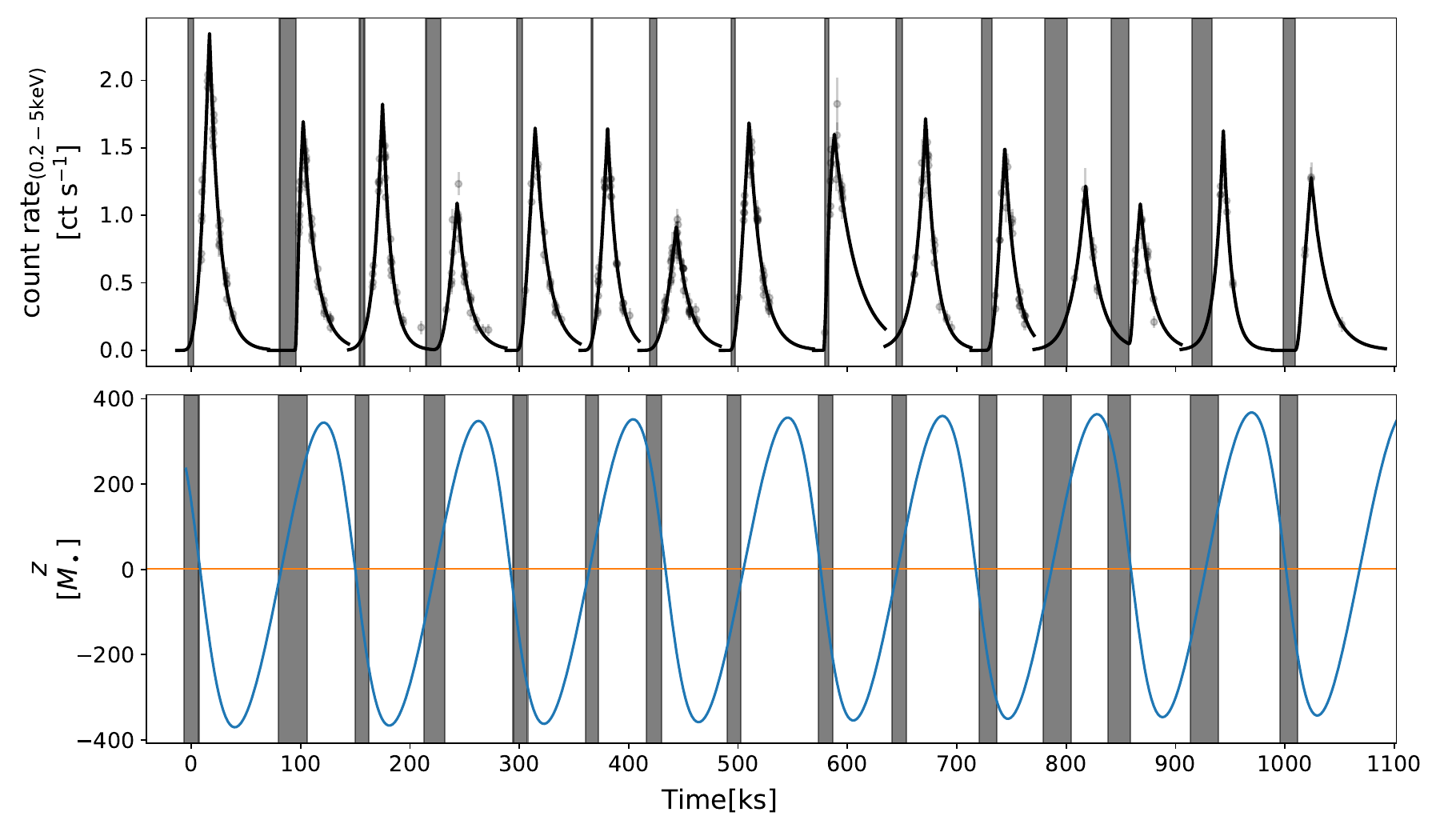}
\caption{\label{fig:eRO1_202108_lc} Same to Fig.~\ref{fig:GSN_069_lc} except for eRO-QPE 1 (2021-08) with best-fit orbital parameters ($p=407 M_\bullet, e=0.24, T_{\rm obt}= 141.5\ {\rm ks}$).}
\end{figure*}


X-ray eruptions from eRO-QPE 1 were firstly reported in Ref.~\cite{Arcodia2021}, along with eRO-QPE 2. Recently, Chakraborty et al. \cite{Chakraborty2024} summarized the 3.5 years of monitoring eRO-QPE 1, during which $\sim100$ flares in total have been detected with NICER.  
Preliminary analysis yields possible signatures of long-term nodal precession of the accretion disk \cite{Chakraborty2024},
which is not included in our vanilla EMRI+disk model. 
To mitigate possible biases introduced by these unmodeled long-term modulations, 
we choose to the fit the EMRI orbital parameters with short-term observations separately.
Going through the whole 3.5 year observations, 
we find  two short-term observations in Aug. 2020 and Aug. 2021 containing a reasonable number of high-quality flares,
which we use to constrain the EMRI orbital parameters.\footnote{Another observation in Feb. 2022 also contains about 10 resolvable flares.
But these flares are overlapping with their neighbors making the flare timing analysis less straightforward, so we did not include them in the EMRI orbital inference.}

The priors used are the same as the GSN 069 ones (Table~\ref{tab:GSN069}) except $\mathbf{\pi}(p)= \mathcal{U}[50, 5000] \ M_\bullet$, $\mathbf{\pi}(T_{\rm obt}) = \mathcal{U}[70, 700]$ ks, $\mathbf{\pi}(\sigma_{\rm sys})=\mathcal{U}[0, 200]$ ks. The central SMBH mass inferred from the stellar velocity dispersion measurement is $\log_{10}(M_\bullet/M_\odot) = 5.78 \pm 0.55$ \cite{Wevers2022}.

In Figs.~\ref{fig:eRO1_202008_lc} and \ref{fig:eRO1_202108_lc}, we show the EMRI orbits of the best-fit flare timing models for 2020-08 and 2021-08. The posterior corner plots of all the model parameters is shown in Figs.~\ref{fig:eRO1_202008_corner} and \ref{fig:eRO1_202108_corner}, where the orbital parameters are constrained as
\be 
\begin{aligned}
    p &= 701^{+2276}_{-549} \ M_\bullet\ , \\ 
    e &= 0.14^{+0.47}_{-0.13} \ , \quad ({\rm 2020-08})\\
    T_{\rm obt} &= 130.4^{+2.2}_{-2.8}\ {\rm ks}\ ,
\end{aligned}
\ee
and 
\be 
\begin{aligned}
    p &= 727^{+2499}_{-562} \ M_\bullet\ , \\ 
    e &= 0.10^{+0.47}_{-0.10} \ , \quad ({\rm 2021-08})\\
    T_{\rm obt} &= 141.7^{+2.1}_{-2.8}\ {\rm ks}\ ,
\end{aligned}
\ee 
at 2-$\sigma$ confidence level, respectively. 

There is clear evolution in the orbital period $T_{\rm obt}$ from Aug. 2020 to Aug. 2021, 
which is likely the result of the disk evolution: nodal precession and alignment.
In the case of a disk aligning on the equator, the observed orbital period $T_{\rm obt, obs}$ inferred from the flare recurrence times is simply the orbital period.
In the case of a unaligned disk with precession period $T_{\rm prec}$, an inclination angle $\iota_{\rm de}$ relative to the equator, 
and an inclination angle $\iota_{\rm ds}$ relative to the SMO orbit, 
the observed orbital period is modified by the disk precession with 
$T_{\rm obt, obs}= T_{\rm obt} (1 \pm T_{\rm obt}/T_{\rm prec} \times \sin\iota_{\rm de}\sin\iota_{\rm ds})$,
where the $\pm$ sign depends on whether the SMO orbit is prograde or retrograde.
Preliminary analysis in \cite{Chakraborty2024} implies a disk precession period $T_{\rm prec}\sim 6$ days, which is sufficient for explaining the $\sim 11$ ks increase in the observed orbital period
with a small disk alignment from Aug. 2020 to Aug. 2021, $\delta T_{\rm obt, obs} = T_{\rm obt}^2/T_{\rm prec}\times \delta(\sin\iota_{\rm de}\sin\iota_{\rm ds})$. 
To verify this speculation, a full Bayesian analysis 
comparing a more complete EMRI+disk model taking the disk precession into consideration with the vanilla model is necessary.

\subsection{eRO-QPE 2}


\begin{figure*}
\includegraphics[scale=0.58]{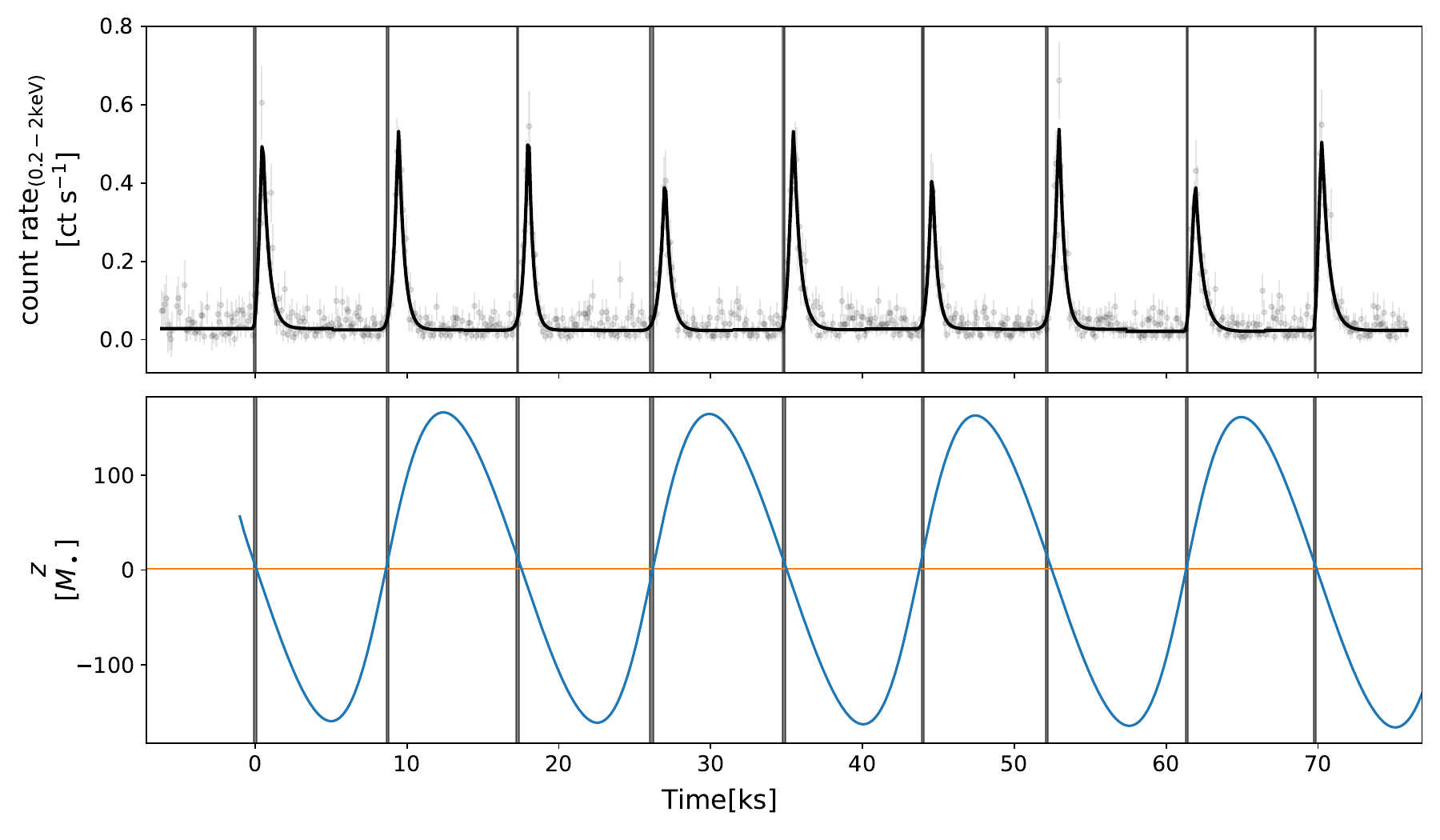}
\caption{\label{fig:eRO2_1_lc} Same to Fig.~\ref{fig:GSN_069_lc} except for eRO-QPE 2 (XMM1) with best-fit orbital parameters ($p=449 M_\bullet, e=0.24, T_{\rm obt}= 17.5\ {\rm ks}$).}
\end{figure*}

\begin{figure*}
\includegraphics[scale=0.58]{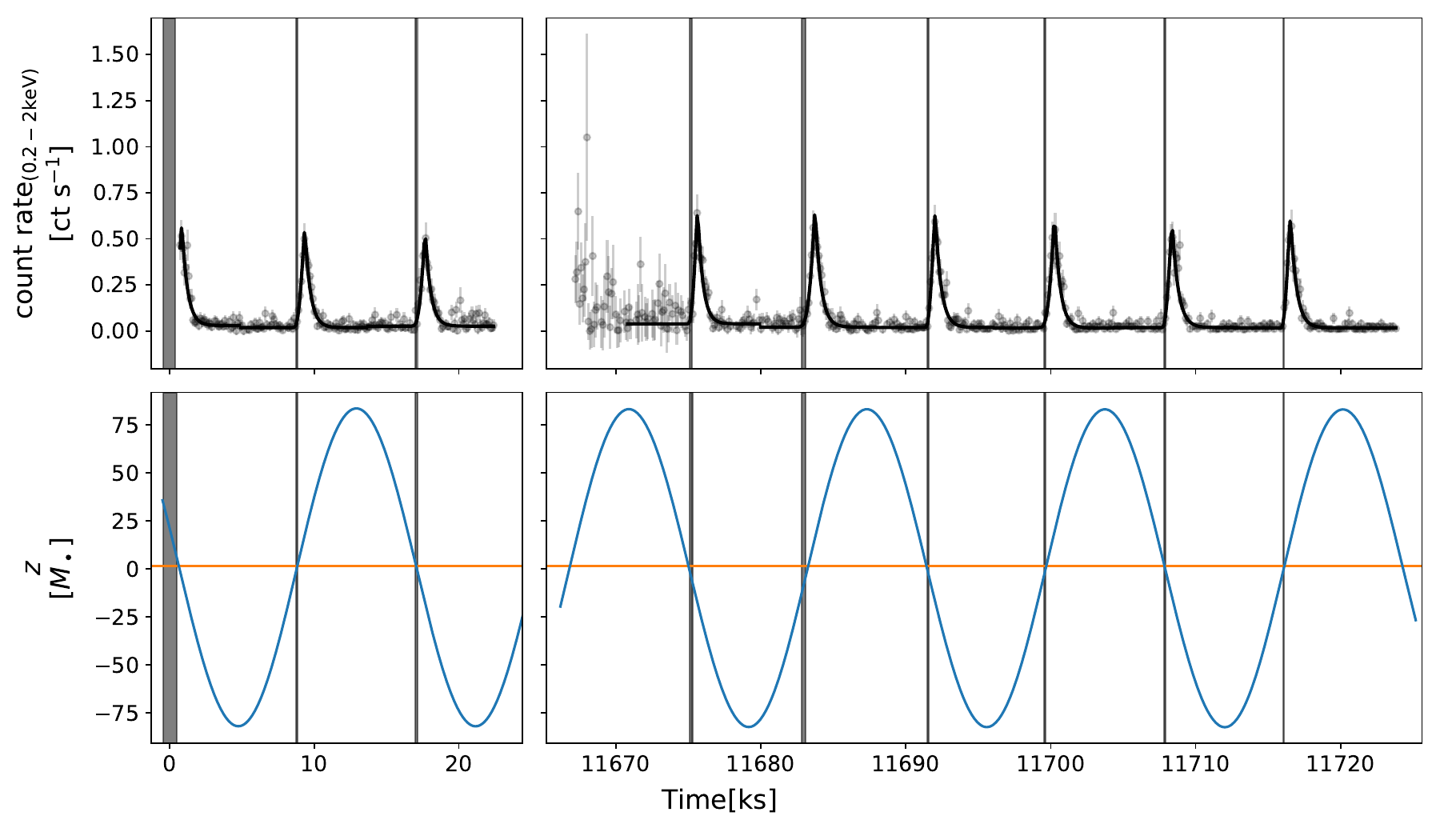}
\caption{\label{fig:eRO2_2_lc} Same to Fig.~\ref{fig:GSN_069_lc} except for eRO-QPE 2 (XMM2 and XMM3) with best-fit orbital parameters ($p=328 M_\bullet, e=0.01, T_{\rm obt}= 16.4\ {\rm ks}$).}
\end{figure*}

Three XMM-Newton observations of eRO-QPE 2 are available.
As we will see later, there is clearly evolution in the orbital period $T_{\rm obt}$ from XMM 1 (2020-08-06) to XMM 2 (2022-02-06) and 3 (2022-06-21),
which is likely the result of SMO-disk interaction. Considering the evolution of the orbital parameters,
we constrain  two sets of EMRI orbital parameters $\mathbf{\Theta}$(XMM 1) and $\mathbf{\Theta}$(XMM 2-3) independently.
The priors used are the same as the GSN 069 ones (Table~\ref{tab:GSN069}) except $\mathbf{\pi}(p)= \mathcal{U}[10, 500] \ M_\bullet$, $\mathbf{\pi}(T_{\rm obt}) = \mathcal{U}[5, 40]$ ks, $\mathbf{\pi}(\sigma_{\rm sys})=\mathcal{U}[0, 3]$ ks for XMM 1 and $\mathbf{\pi}(T_{\rm obt}) = \mathcal{U}[10, 30]$ ks for XMM 2-3.
The central SMBH mass inferred from the stellar velocity dispersion measurement is $\log_{10}(M_\bullet/M_\odot) = 4.96 \pm 0.54$ \cite{Wevers2022}.

In Figs.~\ref{fig:eRO2_1_lc} and \ref{fig:eRO2_2_lc}, we show the EMRI orbits of the best-fit flare
timing models for observations XMM 1 and XMM 2-3, respectively.
The posterior corner plots of all the model parameters are shown in Figs.~\ref{fig:eRO2_1_corner} and \ref{fig:eRO2_2_corner}, 
where the orbital parameters are constrained as
\be 
\begin{aligned}
    p &= 290^{+194}_{-189} \ M_\bullet\ , \\ 
    e &= 0.06^{+0.20}_{-0.05} \ , \quad ({\rm XMM\ 1})\\
    T_{\rm obt} &= 17.5^{+0.1}_{-0.1}\ {\rm ks}\ ,
\end{aligned}
\ee
and 
\be 
\begin{aligned}
    p &= 342^{+146}_{-205} \ M_\bullet\ , \\ 
    e &= 0.01^{+0.05}_{-0.01} \ , \quad ({\rm XMM\ 2-3})\\
    T_{\rm obt} &= 16.4^{+0.2}_{-0.1}\ {\rm ks}\ ,
\end{aligned}
\ee 
at 2-$\sigma$ confidence level, respectively. 
The constraint on the orbital eccentricity in XMM 2-3 turns out to be much tighter than in XMM 1, 
because the SMO  apsidal precession can be much better constrained in XMM 2-3 which spans a much longer time.

In XMM 1, there is a visible alternating long-short pattern in the flare recurrence times,
while the orbital eccentricity turns out to be consistent with zero (see Fig.~\ref{fig:eRO2_1_corner}).
As explained in Sec.~\ref{subsec:GSN069}, 
it is possible to observe such pattern even in the
case of a circular SMO orbit because of different light prop-
agation delays from the two collision locations in one orbital
period to the observer.

From XMM 1 to XMM 2-3, there is a clear decay in the EMRI orbital period $T_{\rm obt}$,
and a likely origin is the extra dissipation due to the collisions between the SMO and  the accretion disk,
\be 
\dot T_{\rm obt} = \frac{dT_{\rm obt}}{dE_{\rm bind}} \dot{E}_{\rm bind} =  \frac{dT_{\rm obt}}{dE_{\rm bind}} \frac{2\delta E}{T_{\rm obt}}
=-3\frac{\delta E }{E_{\rm bind} }\ ,
\ee 
where $\delta E$ is the orbital energy loss per collision and $E_{\rm bind}$ is the binding energy of the SMO.
Assuming a standard $\alpha$ disk model with an accretion rate $\dot M_\bullet$ \cite{SS1973},  
the disk surface density $\Sigma(r)$ and the disk thickness to the mid-plane $H(r)$ in the radiation dominated regime are analytically available \cite{Kocsis2011}.
As shown in Paper I, the sBH orbital change rate due to collisions with the accretion disk is 
\be\label{eq:Tdot}
\begin{aligned}
    \dot T_{\bullet, \rm obt} & \approx  -2\times 10^{-5}\left(\frac{\ln\Lambda}{10}\right)   \\ 
&\times  \alpha_{0.01}^{-1} \dot M_{\bullet,0.1}^{-1} M_{\bullet,5}^{-7/3} T_{\rm obt, 17}^{7/3}
\left( \frac{m_\bullet}{30 M_\odot}\right)  \left(\frac{\sin\iota_{\rm sd}}{0.1}\right)^{-3} ,
\end{aligned}
\ee 
where  $\ln\Lambda= \ln(b_{\rm max}/b_{\rm min})$ is the Coulomb logarithm
with $b_{\rm max/min}$ the maximum/minimum cutoff distance associated to the interaction, 
$\iota_{\rm sd}$ is the angle between the sBH orbital plane
and the disk plane,
and we have defined $T_{\rm obt, 17}:= T_{\rm obt}/17\ {\rm ks}$,
 $\alpha_{0.01}=\alpha/0.01,  \dot M_{\bullet,0.1}=\dot M_\bullet/(0.1 \dot M_{\bullet,\rm Edd})$ 
with $\dot M_{\bullet,\rm Edd}$ the Eddington accretion rate.
The star orbital period change rate is
\be \label{eq:Tdot_star}
\begin{aligned}
    \dot T_{\star,\rm obt} 
    &\approx -5\times 10^{-5} \alpha_{0.01}^{-1} \dot M_{\bullet,0.1}^{-1} M_{\bullet,5}^{-1}  T_{\rm obt, 17}
m_{\star,\odot}^{-1} R_{\star,\odot}^2 \sin\iota_{\rm sd}\ .
\end{aligned}
\ee 
As a result, both a sBH EMRI and a stellar EMRI are  consistent with the observation $\dot T_{\rm obt}\approx -1.7  \times 10^{-5}$
if the central SMBH is on the lower mass end $M_\bullet \sim  10^5 M_\odot$. 

On the other hand, the short orbital period implies the  SMO nature: a main sequence star may not survive on such a tight orbit.
For a star orbiting around a SMBH with orbital radius $r$, the Roche lobe radius $r_{\rm Roche}$ can be estimated  as \cite{Paczynsky1971}
\be 
\frac{r_{\rm Roche}}{r} = 0.46224 \left( \frac{q}{1+q}\right)^{1/3} \approx 0.01  M_{\bullet, 5}^{-1/3}m_{\star,\odot}^{1/3}\ ,
\ee 
where $q:=m_\star/M_\bullet$ is the mass ratio.
In combination with the following two equations of the semi-major axis $A$ ($\approx r$ for the low-eccentricity EMRI in eRO-QPE 2) and the star tidal radius $r_{\rm t}$
\be 
\frac{r}{M_\bullet} = 308 \ T_{\rm obt, 17}^{2/3} M_{\bullet,5}^{-2/3}\ ,
\ee 
and 
\be 
\frac{r_{\rm t}}{M_\bullet} = 217 \ M_{\bullet,5}^{-2/3} R_{\star,\odot} m_{\star,\odot}^{-1/3}\ , 
\ee 
one can find 
\be 
\frac{r}{2 r_{\rm t}}\approx \frac{r_{\rm Roche}}{R_\star} \approx 0.7\   T_{\rm obt, 17}^{2/3}  R_{\star,\odot}^{-1} m_{\star,\odot}^{1/3}\ .
\ee 
To avoid the Roche lobe overflow, the SMO must be either a sBH or a star of very low mass 
$m_\star \lesssim 0.4 M_\odot$, where we have used the stellar mass-radius relation $R_\star \propto m_\star^{0.8}$.

\subsection{Swift J023017}\label{subsec:swift}

\begin{figure*}
\includegraphics[scale=0.58]{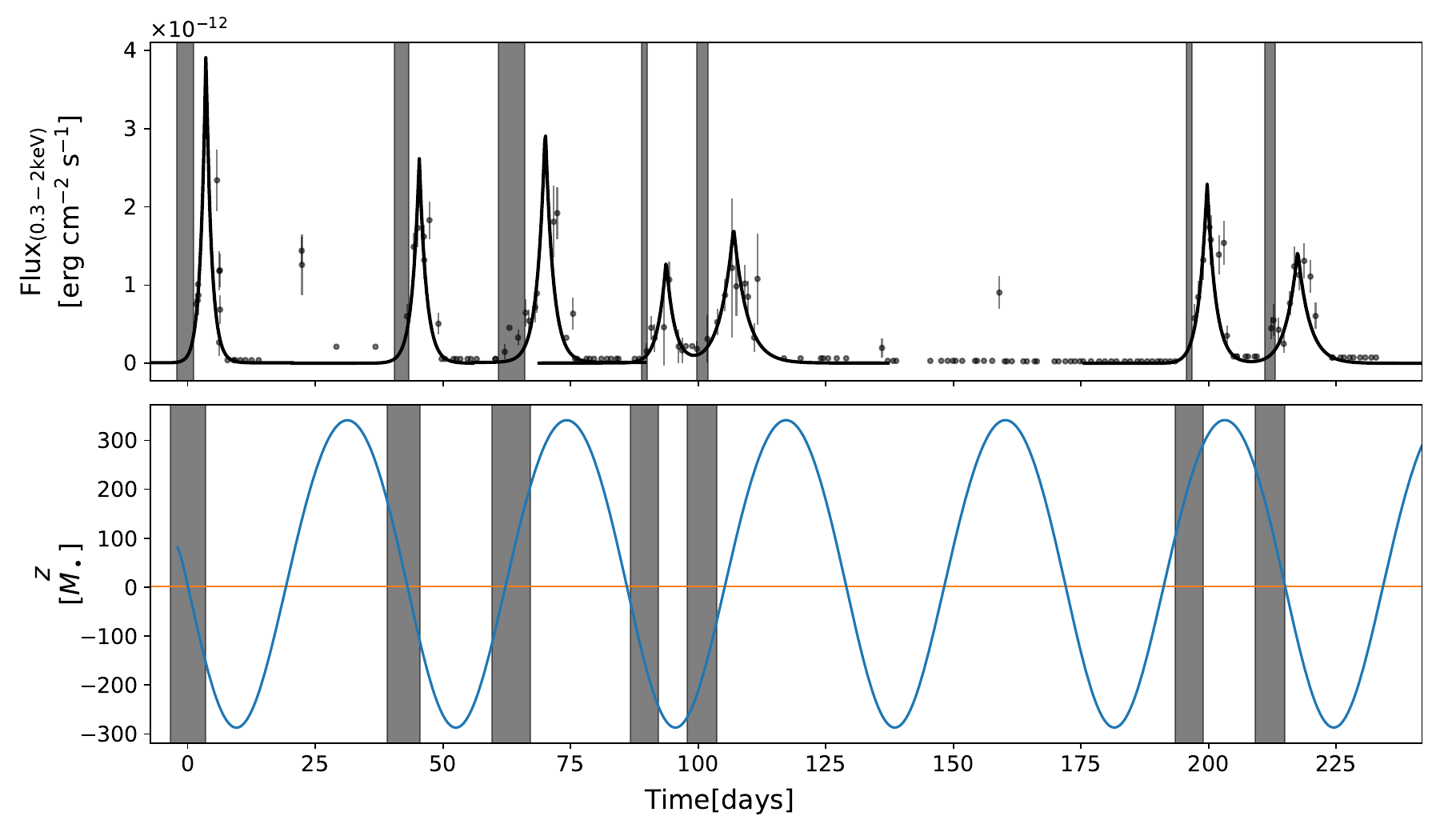}
\caption{\label{fig:Swift_lc} Same to Fig.~\ref{fig:GSN_069_lc} except for Swift J023017 with best-fit orbital parameters ($p=1017 M_\bullet, e=0.09, T_{\rm obt}= 43.0\ {\rm d}$). In the upper panel, we see the best-fit light curves are quite symmetric.}
\end{figure*}

The flare recurrence time in Swift J023017 is about 3 weeks long, which is much longer than in any other QPE sources \cite{Guolo2024,Evans2023}.
Other than the long recurrence time, general properties of Swift J023017 are quite similar to those of other QPE sources,
with the exception that the light curves of  Swift J023017 flares look slightly asymmetric
(slow rise and fast decay, see Table~1 in \cite{Guolo2024} for a good summary). As we will show later in Fig.~\ref{fig:Swift_lc}, the asymmetry is in fact a visual illusion, 
because a detailed light curve fitting shows that these flares are actually quite symmetric, 
showing no clear trend of slow rise and fast decay, similar to other QPEs.
Therefore Swift J023017 should be classified as a QPE source with a long recurrence time.

Both Swift data and NICER data of Swift J023017 are available as shown in Refs.~\cite{Guolo2024,Evans2023}. 
We use the Swift data only for our analysis due to the better data quality and completeness.
The priors used are the same as the GSN 069 ones (Table~\ref{tab:GSN069}) except $\mathbf{\pi}(p)= \mathcal{U}[50, 2000] \ M_\bullet$, $\mathbf{\pi}(T_{\rm obt}) = \mathcal{U}[5, 50]$ days, $\mathbf{\pi}(\sigma_{\rm sys})=\mathcal{U}[0, 100]$ days. The central SMBH mass inferred from the stellar velocity dispersion measurement is $\log_{10}(M_\bullet/M_\odot) = 6.6 \pm 0.4$ \cite{Guolo2024}.

Due to the limited number of flares available, most of the model parameters are not well constrained (see the corner plot in Fig.~\ref{fig:Swift_corner} for details). Fortunately, the two most important orbital parameters $T_{\rm obt}$ and $e$ are
constrained with reasonable uncertainties which as we will show later yield valuable information of its EMRI formation history. 

In Fig.~\ref{fig:Swift_lc}, we show the EMRI orbit of the best-fit flare timing model (with orbital parameters $p=1017 \ M_\bullet, e=0.09, T_{\rm obt}=43.0 \ {\rm days})$ 
along with the starting time of each flare. 
The posterior corner plot of all the model parameters is shown in Fig.~\ref{fig:Swift_corner}, where the orbital parameters are constrained as
\be 
\begin{aligned}
    p &= 963^{+942}_{-651} \ M_\bullet\ , \\ 
    e &= 0.22^{+0.50}_{-0.21} \ , \\
    T_{\rm obt} &= 39.6^{+8.0}_{-9.7}\ {\rm d}\ , 
\end{aligned}
\ee 
at 2-$\sigma$ confidence level, respectively. 

There are in total 7 flares with both the rising and the decay parts resolved, which we used in the EMRI orbital analysis.
From the best-fit orbit, we expect another 4 flares arising from 4 SMO-disk collisions at $t\approx 20, 130, 150, 170$ d, respectively (see Fig.~\ref{fig:Swift_lc}).  The first two are consistent with two vaguely visible flares, while there is no visible flare found around the latter two disk crossing times. The flare disappearance is hard to interpret in the vanilla EMRI+disk model.

Constrained by the exceptionally long recurrence time, the orbital radius of the Swift J023017 EMRI must be exceptionally large (compared with other QPE sources). 
Consequently the accretion disk of  Swift J023017 must be much larger than a normal TDE disk for the SMO-disk collisions to happen. 
This implies that there are other origins of the accretion disks in QPE sources in addition to TDEs.
And optical emission-line diagnostic diagrams of the host galaxy of Swift J023017   
indeed indicates a low-luminosity AGN \cite{Guolo2024}. 
If this is true, there should be two different types of QPEs: some are ignited by a TDE, and some are not.
This speculation could be answered with more QPE discoveries in the coming years.

\section{Summary and Discussion}\label{sec:summary}

QPEs are invaluable in inferring the formation of EMRIs.
In this section, we first infer the possible formation channels of the 5 EMRIs analyzed in this work from their current orbital parameters,
then discuss their implication on future GW detection of EMRIs, and finally discuss what extra information 
regarding the EMRI+disk system can be extracted from long term observations of QPEs.

\subsection{EMRI formation}
In the EMRI+disk framework, we have done the EMRI orbital analysis for 5 QPE sources, where 4 EMRIs are found to be low-eccentricity (consistent with 0) and 1 is mildly eccentric ($e\approx 0.25$).
These SMOs may be  captured by the SMBH via the dry loss-cone channel \cite{Hopman2005,Preto2010,Bar-Or2016,Babak2017,Amaro2018,Broggi2022}, the Hills mechanism (binary disruption) \cite{Miller2005,Raveh2021} or the wet AGN disk channel \cite{Sigl2007,Levin2007,Pan2021prd,Pan2021b,Pan2021,Pan2022,Derdzinski2023,Wang2023,Wang2023b}.
 As shown in Paper I, the EMRI formation history can be inferred from its current orbital parameters. 
 Arcodia et al. \cite{Arcodia:2024efe} compared the formation rates of 
QPEs and TDEs, and found the former is a factor of $\sim 10 (t_{\rm QPE}/1 {\rm yr})$ smaller than the latter,
where $t_{\rm QPE}$ is the average QPE lifetime. In this work, we use the ratio $\sim 10^{-3}$ as a benchmark in inferring the potential formation channels of EMRIs sourcing QPEs assuming a long QPE lifetime $t_{\rm QPE}\sim 10^2$ yr \cite{Linial:2024mdz,Yao:2024rtl}.
 We examine the 5 EMRIs following the discussion of Paper I and 
 we outline the argument as follows (see Section IV. B of Paper I for details).

Considering  a star in a stellar cluster around a SMBH, its motion 
is mainly affected by two processes: 2-body scatterings by other SMOs in the cluster and GW emission.
Now we examine the typical timescales of these two processes, 
which provide convenient guidelines for understanding the star motion in the stellar cluster.
The relaxation timescale of the star at radius $r$ due to 
2-body scatterings is \cite{Spitzer1987}
\be 
t_{\rm rlx}(r) = \frac{0.339}{\ln\Lambda}\frac{\sigma^3(r)}{m_\star^2 n_\star(r)}\ ,
\ee 
where $\sigma(r)$ is the local velocity dispersion ($\approx\sqrt{M_\bullet/r}$ within the influence radius $r_{\rm h}:=M_\bullet/\sigma_\star^2$ of the SMBH),
$n_\star(r)$ is the star number density, and $\ln\Lambda\approx10$ is the  Coulomb logarithm.
On a non-circular orbit, the diffusion timescale in the angular momentum in general is
shorter than the relaxation timescale in the energy as $t_J\approx (1-e^2) t_{\rm rlx}$, i.e., the pericenter distance $r_{\rm p}$
is easier to be changed than the semimajor axis $A$ by 2-body scatterings.
And the energy dissipation timescale of the star $t_{\rm GW}$ due to GW emission is given in Ref.~\cite{Peters1964}.
Assuming the Bahcall-Wolf (BW) density profile $n_\star(r) \propto r^{-7/4}$ \cite{Bahcall1976}, 
one can find the ratio of the two timescales \cite{Linial2023b}
\be \label{eq:tratio}
\frac{t_J}{t_{\rm GW}} \approx \left(\frac{r_{\rm p}}{M_\bullet}\right)^{-5/2}\left(\frac{A}{r_{\rm h}}\right)^{-5/4}\ ,
\ee  
where $r_{\rm p}$ and $A$ are the star pericenter distance and the semi-major axis, respectively.

\begin{figure*}
\includegraphics[scale=0.7]{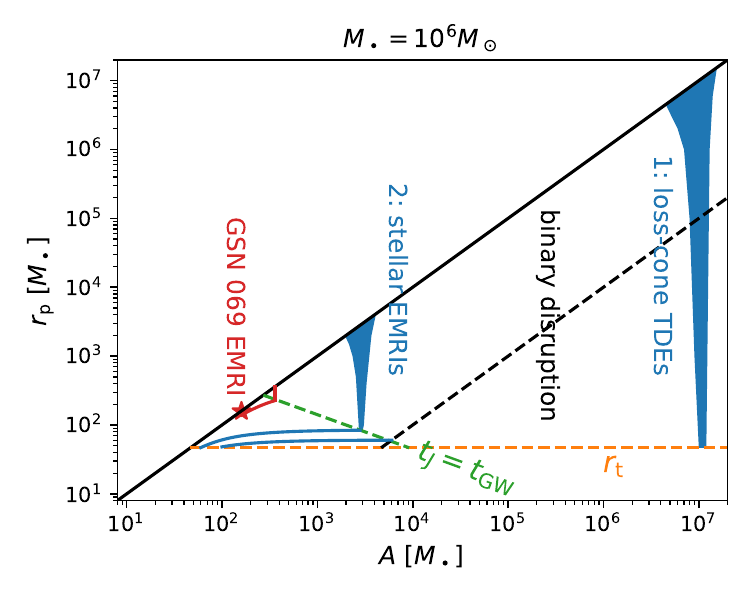}
\includegraphics[scale=0.7]{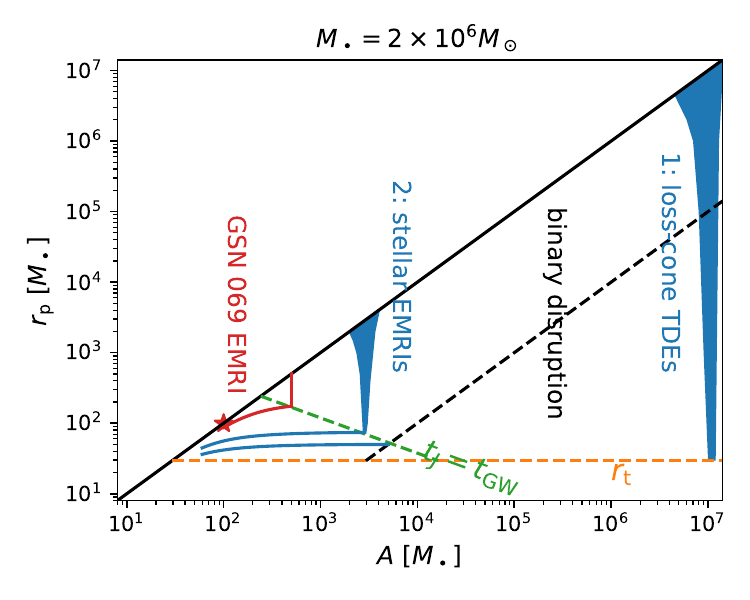}
\caption{Same to  Paper I, we use the $r_{\rm p}-A$ (pericenter distance-semimajor axis)  diagram to examine the possibility of forming EMRIs in QPE sources via the loss-cone channel or the Hills mechanism. In the loss-cone channel, stars generally end in two fates: (1) loss-cone TDEs, most of them are
dominated by 2-body scatterings ($t_J < t_{\rm GW}$) and get tidal disrupted 
by the SMBH when occasionally scattered into a low-angular momentum orbit with $r_{\rm p} < r_{\rm t}$, 
and (2) stellar EMRIs, a small fraction of them are scattered into the GW emission dominated regime ($t_J > t_{\rm GW}$) gradually
circularizing and losing mass via partial TDEs. 
In the Hills mechanism, the bounded star after a
binary disruption is highly eccentric  with eccentricity $e\approx 0.99$ (the black dashed lines) and the star faces the same two fates, a loss-cone TDE or a stellar EMRI.
The markers denote EMRIs in the QPE sources with the errorbars denoting the EMRI orbital eccentricities at $2-\sigma$ confidence level.\\
\emph{Left Panel}: Taking GSN 069 as an example assuming the central SMBH mass $M_\bullet=10^6 M_\bullet$,
stars residing in a tiny phase space ($r_{\rm p}\lesssim 360 M_\bullet \ {\rm or}, e\approx 0$) can possibly become GSN 069  like EMRIs. The bounded stars after a binary disruption either becomes a TDE or an eccentric stellar EMRIs, neither of which become GSN 069 like EMRIs. \\
\emph{Right Panel}: same to the left except assuming a different SMBH mass. }
\label{fig:no_lc}
\end{figure*}

Making use of  $r_{\rm p}-A$ phase diagram \cite{Linial2023b}, we first consider the possibility of forming  GSN 069 like EMRIs
in the dry loss-cone channel or via the Hills mechanism (Fig.~\ref{fig:no_lc}). 
Here GSN 069 like EMRIs are  EMRIs residing in the GW dominated regime ($t_J> t_{\rm GW}$) with orbital parameters $(T_{\rm obt} ,e)$ that are/were/will be confidently consistent with that of the GSN 069 EMRI,
i.e., $
(T_{\rm obt} ,e) \xrightarrow[]{\text{GW}} (T_{\rm obt,  GW}> 63 \ {\rm ks}, e_{\rm GW}< 0.13)
$.
In the phase diagram of the left panel, we consider a SMBH mass $M_\bullet=10^6 M_\odot$ ($A=160 M_\bullet$)
it is clear that most stars reside in the scattering dominated regime with $t_J < t_{\rm GW}$ and  
get tidally disrupted when occasionally scattered into
a low-angular momentum orbit with $r_{\rm p} < r_{\rm t}:=(M_\bullet/m_\star)^{1/3} R_\star$.
A small fraction of them are scattered into the GW emission dominated corner with $t_J > t_{\rm GW}$ (dubbed as stellar EMRIs),
gradually circularizing and losing mass via partial TDEs.
The TDE rate and stellar EMRI formation rate can be estimated as 
\be 
R_{\star}(< r) \approx \frac{N_\star(<r)}{t_{\rm rlx} (r)} \propto r\ ,
\ee 
where $N_\star(<r)$ is the total number of stars within radius $r$.
From the phase diagram, it is clear that only stars in the range of $r < 360 M_\bullet, e\approx 0$ can
possibly become  GSN 069 like EMRIs (the result slightly differs from in Paper I due to mild changes in the model parameter constraints). 
\be 
f_{\rm QPE/TDE}^{\rm GSN\ 069}(M_\bullet=10^6M_\odot)=\frac{R_\star(r < 360 M_\bullet)}{R_\star(r < r_{\rm h})}\approx 1.8\times 10^{-5}\ .
\ee 
Since the central SMBH mass is not perfectly known, we also consider other possible values, e.g., $M_\bullet=2\times 10^6 M_\odot$ ($A=100 M_\bullet$, the lower edge of the 2-$\sigma$ confidence region) in the right panel of Fig.~\ref{fig:no_lc}.
In this more massive SMBH case, the semi-latus rectum $p$ of the SMO is of lower value and 
the GW dominated regime where $t_J > t_{\rm GW}$ expands to a larger parameter space, therefore the GSN 069 EMRI has gone through
a longer circularization history.  We calculate the ratio $f_{\rm QPE/TDE}^{\rm GSN\ 069}(M_\bullet)$ for general SMBH mass $M_\bullet$,
and find the dependence can be approximately fitted as 
\be 
f_{\rm QPE/TDE}^{\rm GSN\ 069}(M_\bullet)\approx 1.8\times 10^{-5}\times \left(\frac{M_\bullet}{10^6 M_\odot}\right)^{1.0}\ .
\ee 
Therefore in the reasonable mass range of GSN 069 SMBH, the ratio $f_{\rm QPE/TDE}^{\rm GSN\ 069}(M_\bullet)$ is far below the benchmark ratio $10^{-3}$.
The ratio become even lower if considering the mass-segregation effect which is expected to suppress the light component (stars) density 
at small radii \cite{Hopman2005,Preto2010,Bar-Or2016,Babak2017,Amaro2018,Broggi2022}.
Therefore the stellar EMRI in GSN 069  was unlikely born in the dry loss-cone channel.

\begin{figure*}
\includegraphics[scale=0.7]{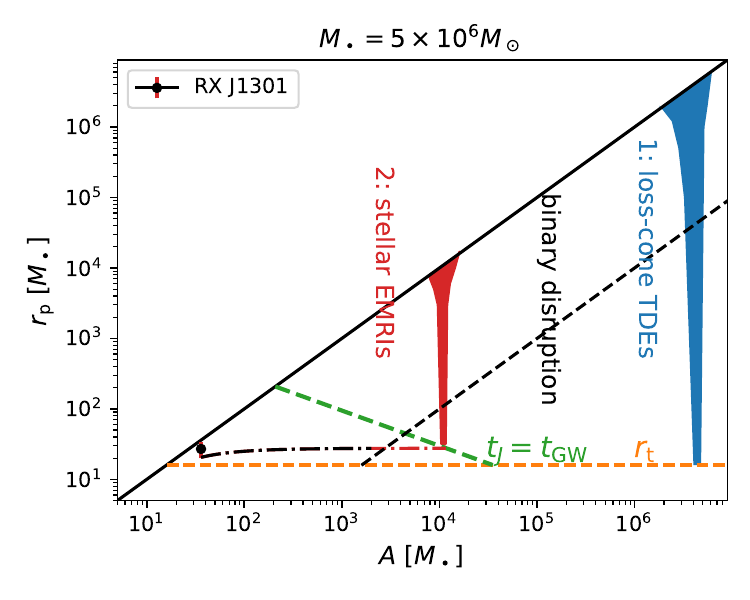}
\includegraphics[scale=0.7]{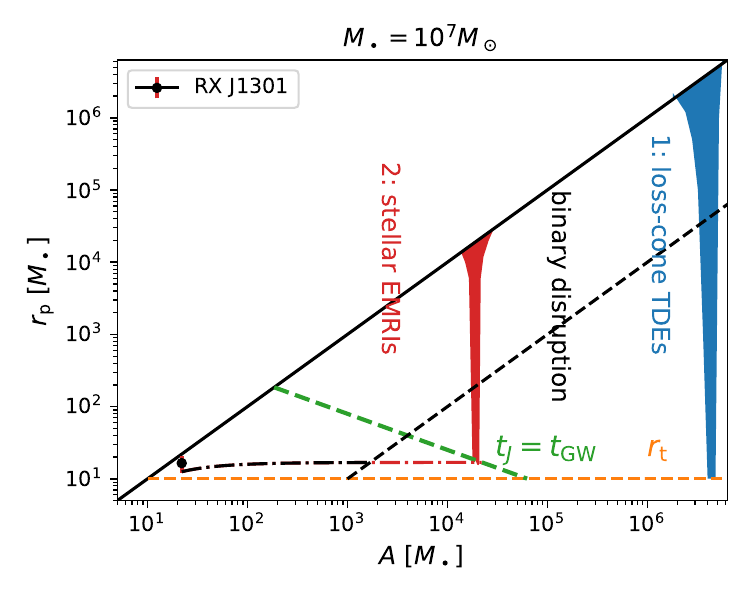}
\caption{Same to Fig.~\ref{fig:no_lc} except for RX J1301 like EMRIs, which can be produced in both the dry loss-cone channel (red dotted-dashed lines) and the Hills mechanism (black dotted-dashed lines). }
\label{fig:yes_lc}
\end{figure*}

In the same way, we find 
the ratio of  RX J1301 like EMRIs [EMRIs residing in the GW emission dominated regime with orbital parameters $(T_{\rm obt} ,e) \xrightarrow[]{\text{GW}} (T_{\rm obt,  GW}> 31 \ {\rm ks}, e_{\rm GW}< 0.43)$]
 is approximately
\be \label{eq:f_RXJ1301}
f_{\rm QPE/TDE}^{\rm RX\ J1301}(M_\bullet)\approx 1.3\times 10^{-3}\times \left(\frac{M_\bullet}{5\times 10^6 M_\odot}\right)^{1.4}\ ,
\ee 
the ratio of  eRO-QPE 1 like EMRIs [EMRIs residing in the GW emission dominated regime with orbital parameters $(T_{\rm obt} ,e) \xrightarrow[]{\text{GW}} (T_{\rm obt,  GW}> 139 \ {\rm ks}, e_{\rm GW}< 0.57)$] is approximately
\be 
f_{\rm QPE/TDE}^{\rm eRO-QPE 1}(M_\bullet)\approx 5.5\times 10^{-5}\times \left(\frac{M_\bullet}{10^6 M_\odot}\right)^{1.25}\ ,
\ee 
the ratio of  eRO-QPE 2 like EMRIs [EMRIs residing in the GW emission dominated regime with orbital parameters $(T_{\rm obt} ,e) \xrightarrow[]{\text{GW}} (T_{\rm obt,  GW}> 16.3 \ {\rm ks}, e_{\rm GW}< 0.06)$]
 is approximately
\be 
f_{\rm QPE/TDE}^{\rm eRO-QPE 2}(M_\bullet)\approx 0.7\times 10^{-5}\times \left(\frac{M_\bullet}{10^5 M_\odot}\right)^{0.5}\ ,
\ee 
and the ratio of  Swift J02317 like EMRIs [EMRIs residing in the GW emission dominated regime with orbital parameters $(T_{\rm obt} ,e) \xrightarrow[]{\text{GW}} (T_{\rm obt,  GW}> 30 \ {\rm d}, e_{\rm GW}< 0.72)$] is approximately
\be 
f_{\rm QPE/TDE}^{\rm Swift\ J02317}(M_\bullet)\approx 8\times 10^{-5}\times \left(\frac{M_\bullet}{10^7 M_\odot}\right)^{-0.16}\ .
\ee 
Among the above relations, the negative power index of $f_{\rm QPE/TDE}^{\rm Swift\ J02317}(M_\bullet)$ is an exception. 
The reason is simple. The GW emission dominated regime is defined by two boundaries, a right boundary $t_J = t_{\rm GW}$
and a left boundary $e=0$. The majority of QPE EMRIs are found around the $e=0$ boundary, the destination of long inspirals, while the  Swift J02317 EMRI is found around the  $t_J = t_{\rm GW}$ boundary where the inspiral is about to start or has recently started. 

Based on the above rate estimates, 4 of the QPE EMRIs were unlikely born in the dry loss-cone channel, and RX J1301 EMRI is an exception with relatively high orbital eccentricity which is consistent with the dry channel prediction.

\bigskip 

Hills mechanism has been proposed as an alternative efficient EMRI formation channel \cite{Miller2005,Raveh2021}. 
Considering a binary star with total mass $2m_\star$ and an initial binary separation $A_{\rm b}$ orbiting around the SMBH,
the binary disruption occurs at $r_{\rm t, b} \approx (M_\bullet/2m_\star)^{1/3} A_{\rm b} $.
After the binary disruption, in general one star is captured by the SMBH and the other is ejected.
The pericenter distance of the bounded star is identified as  $r_{\rm p} = r_{\rm t, b}$, the lower limit of which is set by contact binaries with $A_{\rm b}=2R_\star$, 
therefore $ r_{\rm p}= r_{\rm t, b} \geq (M_\bullet/2m_\star)^{1/3} 2R_\star \approx r_{\rm t}$, i.e., the lower limit of the binary disruption radius is the single star disruption radius.
The (specific) orbital energy $-E$ of the bounded star is determined by the tidal energy  
\be 
-E \lesssim \frac{M_\bullet}{r_{\rm t,b}} \frac{A_{\rm b}}{r_{\rm t,b}}\ ,
\ee 
therefore its semi-major axis and orbital eccentricity turns out be 
\be 
A \gtrsim r_{\rm t,b} \frac{r_{\rm t,b}}{A_{\rm b}}\ ,
\ee 
and 
\be 
e = 1-\frac{r_{\rm p}}{A }\approx 1-\frac{A_{\rm b}}{r_{\rm t,b}}\gtrsim 1-\left(\frac{2m_\star}{M_\bullet}\right)^{1/3}\ .
\ee 
In the following discussion, we take  $e = 0.99$ as the fiducial value (the back dashed lines in Figs.~\ref{fig:no_lc} and ~\ref{fig:yes_lc}). The bounded star faces the same two fates: either being scattered to a low-angular momentum orbit and gets tidally disrupted, or being scattered to the GW emission dominated regime and becomes a (highly eccentric) stellar EMRI. Neither of them become GSN 069 like EMRIs as shown in Fig.~\ref{fig:no_lc}. Following the same argument, one can show that eRO-QPE 1, eRO-QPE 2,  or Swift 02317 like EMRIs are not the result of binary disruptions (Hills mechanism) either.
RX J1301 is again an exception. As shown in the left panel of Fig.~\ref{fig:yes_lc},
the bounded star from a binary disruption with $A\in (2.5\times 10^3, 1.2\times 10^4) M_\bullet$ 
can become a RX J1301 like EMRI. 

Now we turn to estimate the fraction of binary disruptions contributing to RX J1301 like EMRIs.
For a  nuclear stellar cluster consisting of a central SMBH and surrounding stars (single stars and star binaries),
the binaries relevant are in the empty loss cone regime where the binary disruption rate is independent of the size of the loss cone and only depends on the relaxation timescale of the system \cite{Perets2007}. As a result, the
radial distribution of bounded stars follows the binary separation distribution \cite{Perets2010}.
Assuming a loguniform distribution of the binary separation, 
the semimajor axis $A$ of the bounded star after a binary disruption also follows a loguniform distribution,
$N(<A)\propto \log(A)$ \cite{Perets2010},
then we obtain the fraction 
\be 
\frac{R_{\star, {\rm Hills}}^{\rm RX \ J1301}}{R_{\star, {\rm Hills}}^{\rm TDE}} \approx \frac{\log(1.2\times 10^4 M_\bullet/2.5\times 10^3 M_\bullet)}{\log[r_{\rm h}/(r_{\rm t}/(1-e))]} \approx 0.18\ ,
\ee 
where $r_{\rm t}/(1-e)\approx 1.6\times 10^3 M_\bullet$ is the lower limit coming from contact binaries.
Therefore $\sim 18\%$ of bounded stars from binary disruptions become RX J1301 like EMRIs assuming $M_\bullet=5\times 10^6 M_\odot$.
In the mass range of RX J1301 EMRI, we find the ratio can be approximately fitted as 
\be 
\frac{R_{\star, {\rm Hills}}^{\rm RX \ J1301}}{R_{\star, {\rm Hills}}^{\rm TDE}}  \approx 0.18\times \left(\frac{M_\bullet}{5\times 10^6M_\odot} \right)^{0.6}\ .
\ee 
Therefore RX J1301 like EMRIs could also be from the Hills mechanism.
The relative importance of the dry channel and the Hills mechanism then depends on the uncertain fraction of stars in binaries.
As long as the Hills mechanism contributes more than $1\%$ TDEs, it is expected to dominate the formation of RX J1301 like EMRIs.
One subtlety we did not consider here is that the stability condition [Eq.~(\ref{eq:RX_stable})] requires subsolar mass stars. 

On the other hand, if there are plenty of massive perturbers in the nuclear stellar cluster boosting the binary disruption 
rate by orders of magnitude as considered in \cite{Perets2010}, binaries that are relevant to QPE EMRIs are mostly in the 
full loss cone regime where the binary
disruption rate is proportional to the size of the loss cone.
Assuming a loguniform distribution of the binary separation,
the semimajor axis $A$ of the bounded star after a binary
disruption follows a linear distribution, $N(<A)\propto A$ \cite{Perets2010},
then we obtain the fraction 
\be 
\frac{R_{\star, {\rm Hills}}^{\rm RX \ J1301}}{R_{\star, {\rm Hills}}^{\rm TDE}} \approx \frac{1.2\times 10^4 M_\bullet-2.5\times 10^3 M_\bullet}{r_{\rm h} - (r_{\rm t}/(1-e))} \approx 10^{-3}\  ,
\ee 
for $M_\bullet=5\times10^6 M_\odot$.
In the mass range of RX J1301 EMRI, we find the fraction can be approximately fitted as 
\be 
\frac{R_{\star, {\rm Hills}}^{\rm RX \ J1301}}{R_{\star, {\rm Hills}}^{\rm TDE}}  \approx 10^{-3}\times \left(\frac{M_\bullet}{5\times 10^6M_\odot} \right)^{1.6}\ ,
\ee 
which is comparable to the fraction in the dry channel (Eq.~[\ref{eq:f_RXJ1301}]).

\begin{figure*}
\includegraphics[scale=0.3]{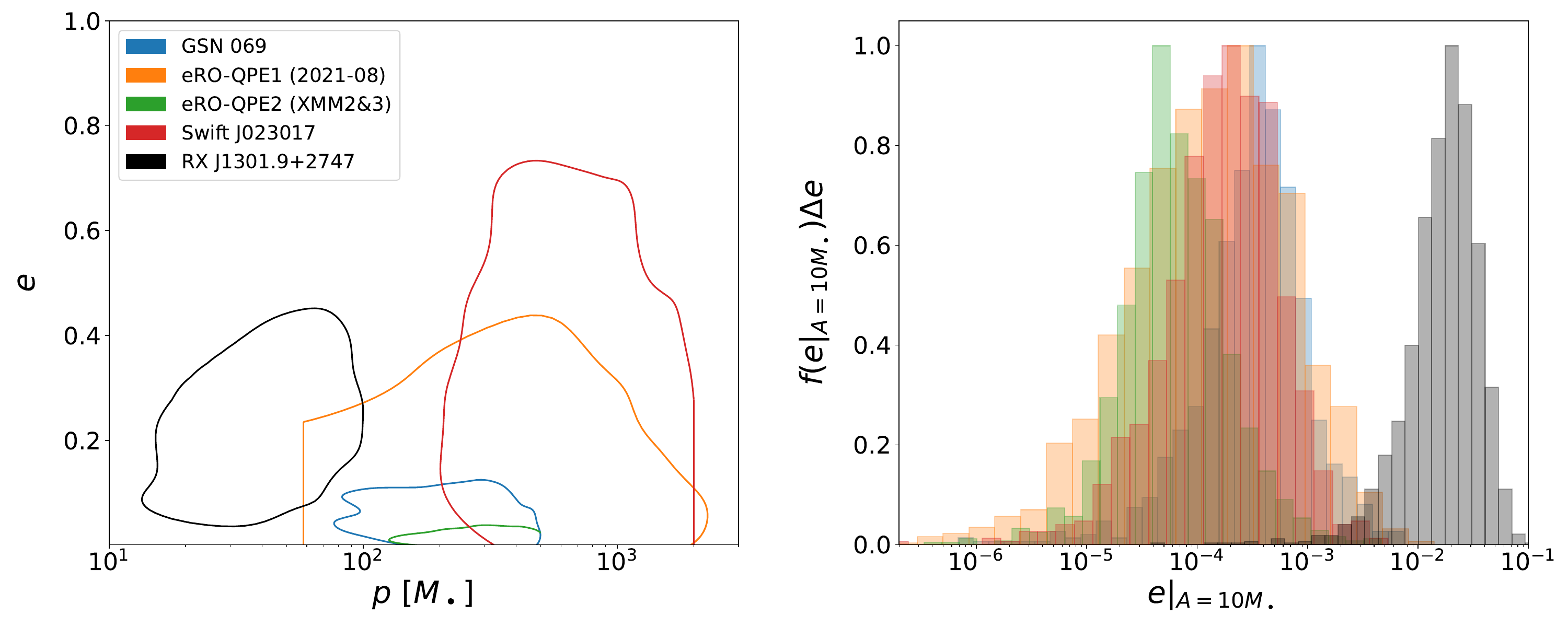}
\caption{Left panel: the constraints of EMRI orbital parameters $e-p$ of the 5 QPE sources analyzed, where the contours enclose parameter space of 2-$\sigma$ confidence level. Right panel: the (unnormalized) probability distributions of the orbital eccentricities $e |_{A = 10 M_\bullet}$ as the EMRIs entering the sensitivity band of spaceborne GW detectors $f(e|_{A = 10 M_\bullet}) \Delta e$, where $\Delta e$ is the size of the bin centered at $e$.}
\label{fig:e-p_statistics}
\end{figure*}

\bigskip

In the wet EMRI formation channel, a SMO orbiting around
an accreting SMBH can be captured to the accretion disk 
driven by interactions (dynamical friction and density waves) with the
accretion disk. After captured onto the disk, the SMO migrates
inward driven by the density waves and gravitational wave
emission. 
The distribution of SMOs in the radial direction is expected to peak at $O(10^2) M_\bullet$ where 
the two driving forces becomes comparable and the migration velocity minimizes
(see Fig.~6 in Paper I for details). 
The orbital eccentricity is expected to be damped
by the density waves to $e\sim h$, where $h:=H/r$ is the disk aspect ratio in the regime by the migration is dominated by density waves.
The 4 low-eccentricity EMRIs in QPE sources GSN 069, eRO-QPE 1, 2 and Swift J023017
are consistent with the prediction of the wet channel. 

In the EMRI+disk framework, Swift J023017 EMRI is an exception among the 4 QPE sources with low-eccentricity EMRIs.
Observations of Swift J023017 favor a low-luminosity AGN disk instead of a TDE disk (see Section~\ref{subsec:swift}).
One may expect no flare generation if the EMRI was formed in the wet channel and has been aligned with the 
AGN disk since then. This is true only if the AGN accretion is coherent in a long timescale, i.e.,
the gas is fed from a fixed direction in a time span longer than the merger timescale driven by GW emissions.
In fact, the gas is likely fed from random directions varying from one accretion episode to another,
therefore the EMRI formed in a previous AGN phase is in general misaligned with the current AGN accretion disk.

\bigskip

If the EMRIs sourcing QPEs are a representative sample of sBH EMRIs, we can infer their eccentricities in the sensitively band of spaceborne GW detectors, say $e|_{A = 10 M_\bullet}$. In the left panel of Fig.~\ref{fig:e-p_statistics}, we summarize the EMRI orbital parameters $e-p$ inferred from the 5 QPE sources, and  in the right panel we show the projected eccentricity distribution $f(e|_{A = 10 M_\bullet})$ if they are sBH EMRIs and finally plunge into the SMBHs. From both panels, two distinct EMRI populations are identified: 
a low-eccentricity population that is consistent with the prediction of the wet EMRI formation channel and 
a more eccentric population that is consistent with the predictions of both the dry channel and the Hills mechanism.
With an increasing number of QPE sources to be discovered, 
the population properties can be further refined, e.g., 
two eccentric EMRI populations with different eccentricity distributions 
expected from the dry loss-cone channel and the Hills mechanism 
may be identified.

To summarize, QPEs are invaluable in inferring the formation of EMRIs.
In this subsection, we have examined the formation of the 5 EMRIs: 
4 of them should be born in the wet channel, 
and 1 is consistent with the predictions of both the dry channel and the Hills mechanism.
More QPE discoveries in the coming years will promisingly provide detailed answers to the open questions regarding EMRI formation,
and provide reasonable astrophysical priors on the EMRI orbital parameters for future spaceborne GW detection.
All these population properties inferred from QPEs could be tested with future spaceborne GW detection of EMRIs in return.

\subsection{Future work}

In the vanilla EMRI+disk model, we assume a SMO  moving along the geodesic and a steady disk lying on the equator.
With these two assumptions, the EMRI orbital period is constant and is equal to the intervals of flares $T_{\rm long}+T_{\rm short}=T_{\rm obt}$. As noticed in some QPE sources (eRO-QPE 1 and 2) , these assumptions may be violated in the long run.
Considering the EMRI orbital energy dissipation as interacting with the disk, the orbital period is no longer a constant.
The orbital period decay rate $\dot T_{\rm obt}$ can be used for constraining the disk surface density and the SMO mass,
consequently inferring the nature of the SMO, a star or a sBH.
In the case of an unaligned disk around a spinning BH, the disk precession plays a role interpreting the flare timing:
the observed orbital period is different from the true one,
$T_{\rm obt, obs}:=T_{\rm long}+T_{\rm short} = T_{\rm obt}(1 \pm T_{\rm obt}/T_{\rm prec} \times \sin\iota_{\rm de}\sin\iota_{\rm ds})$.
The disk alignment (i.e., the decay of the disk inclination angle $\iota_{\rm de}$) will also contribute to the evolution of the observed orbital period $T_{\rm obt, obs}$.

Without incorporating these long term evolution in the vanilla EMRI+disk model,
we can only fit short term observations, e.g., fitting the XMM 1 and XMM 2-3 observations of eRO-QPE 2 separately. 
A more complete model taking these long term evolution into account can be used to globally fit all  QPE observations with large intervals.
We expect the global fitting to not only largely narrow down the orbital parameter constraints, but also extract new information of the EMRI+disk system. This is what we plan to investigate in detail in a follow-up work.

\acknowledgments
We thank the referee for providing valuable and detailed comments.
We thank Joheen Chakraborty and Michal Zajaček for discussions and kindly sharing their data. We also thank JiaLai Kang for helpful discussions. Lei Huang thanks for the support by National Natural Science Foundation of China (11933007,12325302), Key Research Program of Frontier Sciences, CAS (ZDBS-LY-SLH011),  Shanghai Pilot Program for Basic Research-Chinese Academy of Science, Shanghai Branch (JCYJ-SHFY-2021-013). This work has been supported by the National Key R\&D Program of China (No.2022YFA1603104), and Key Research Program of Frontier Sciences, CAS (Grant No.QYZDJ-SSW-SLH057).

This paper made use of data from XMM-Newton, an ESA science mission with instruments and contributions directly funded
by the ESA Member States and NASA.

\appendix* 
\section{Corner plots}\label{sec:app}

In the Appendix, we show the posterior corner plot of flare timing model parameters for each QPE observation analyzed in the main text.

\begin{figure*}
\includegraphics[scale=0.33]{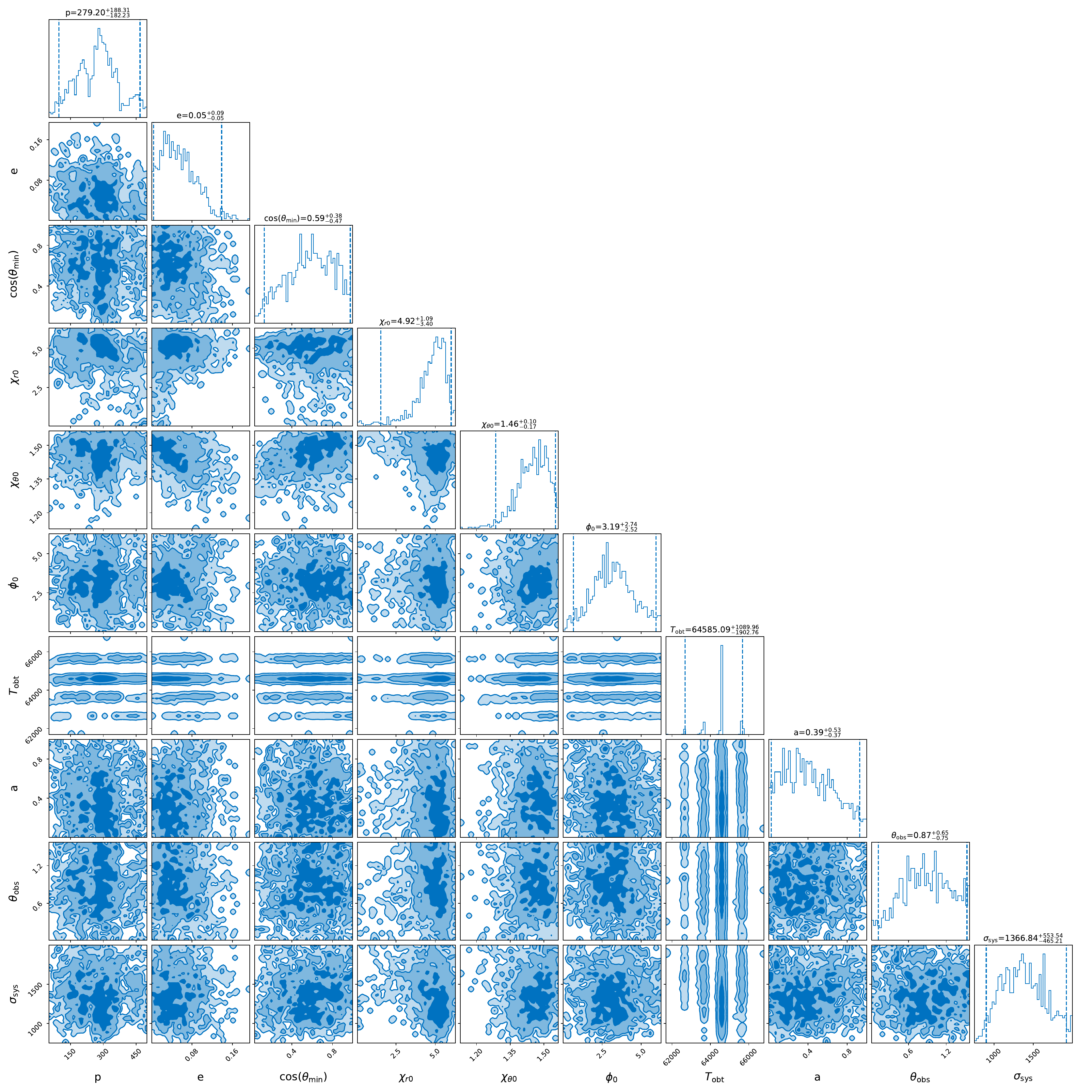}
\caption{\label{fig:GSN_069_corner} The posterior corner plot of the GSN 069 flare timing model parameters: $p [M_\bullet], e, 
\cos\theta_{\rm min}, \chi_{r0}, \chi_{\theta0}, \phi_0,
T_{\rm obt} [{\rm sec}], a, \theta_{\rm obs}, \sigma_{\rm sys} [{\rm sec}]$, where each pair of vertical lines denotes the 2-$\sigma$ confidence level.  }
\end{figure*}

\begin{figure*}
\includegraphics[scale=0.33]{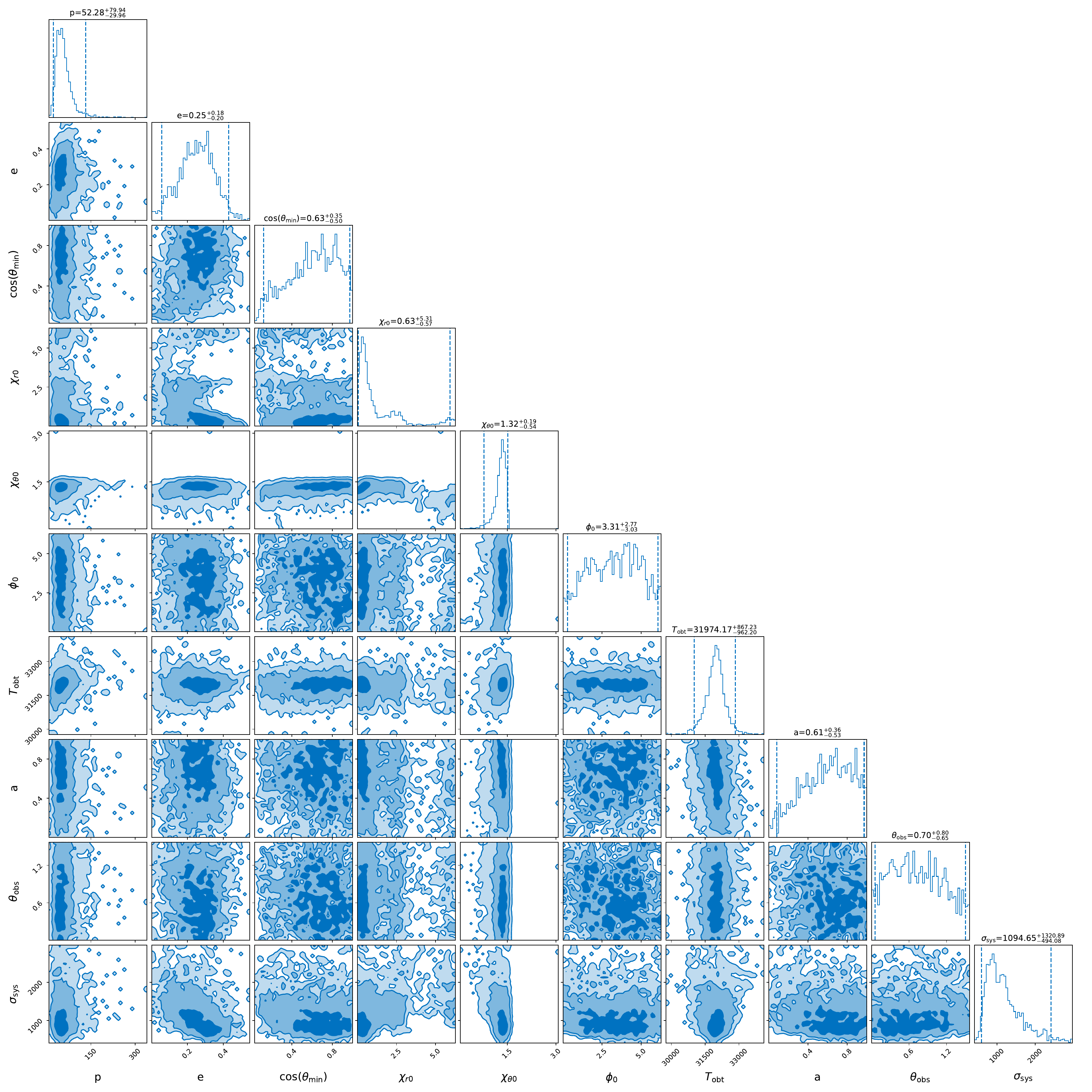}
\caption{\label{fig:RX_corner} Same to Fig.~\ref{fig:GSN_069_corner} except for RX J1301.9+2747. }
\end{figure*}

\begin{figure*}
\includegraphics[scale=0.33]{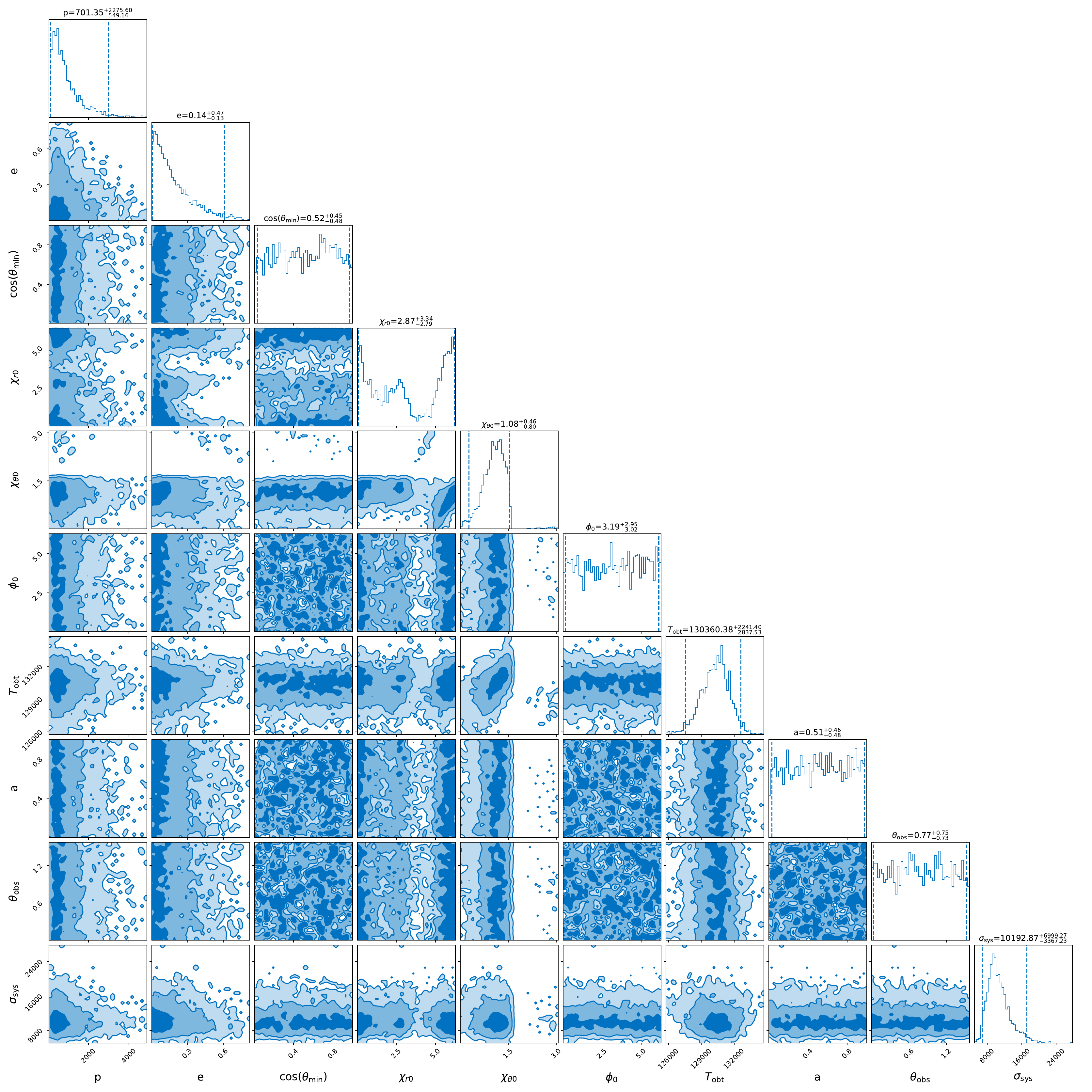}
\caption{\label{fig:eRO1_202008_corner} Same to Fig.~\ref{fig:GSN_069_corner} except for eRO-QPE 1 in 2020-08. }
\end{figure*}

\begin{figure*}
\includegraphics[scale=0.33]{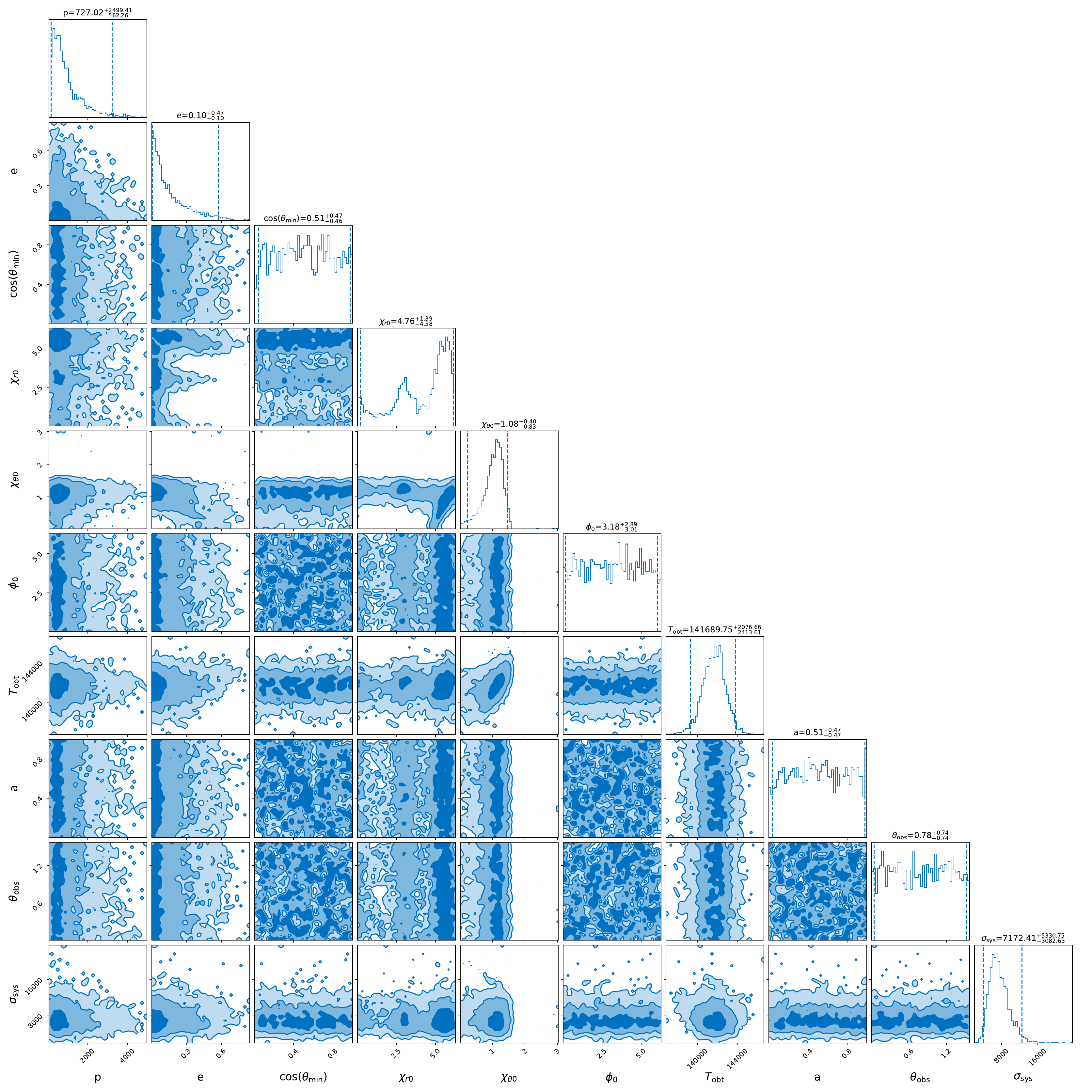}
\caption{\label{fig:eRO1_202108_corner} Same to Fig.~\ref{fig:GSN_069_corner} except for eRO-QPE 1 in 2021-08. }
\end{figure*}


\begin{figure*}
\includegraphics[scale=0.33]{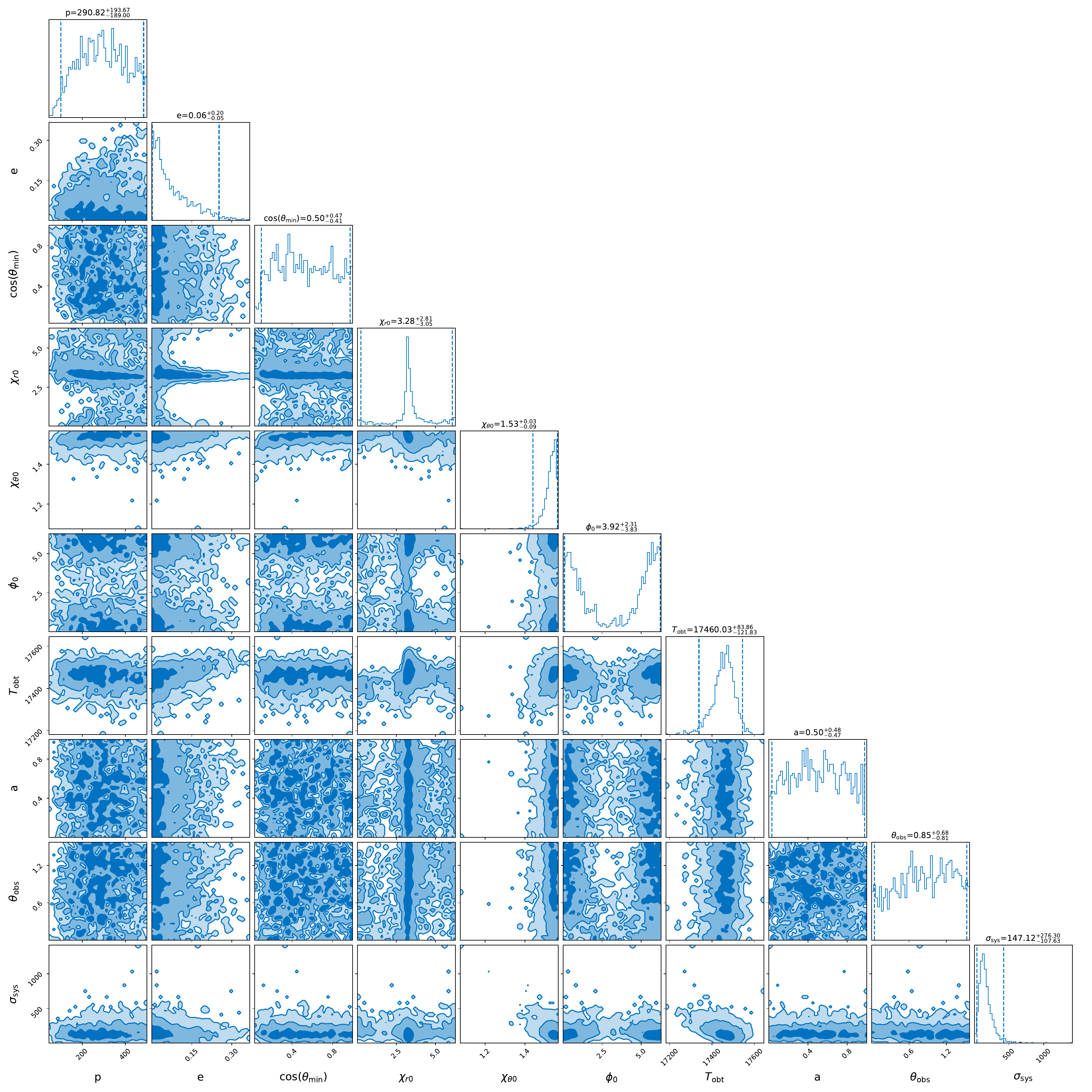}
\caption{\label{fig:eRO2_1_corner} Same to Fig.~\ref{fig:GSN_069_corner} except for eRO-QPE 2 (XMM1). }
\end{figure*}

\begin{figure*}
\includegraphics[scale=0.33]{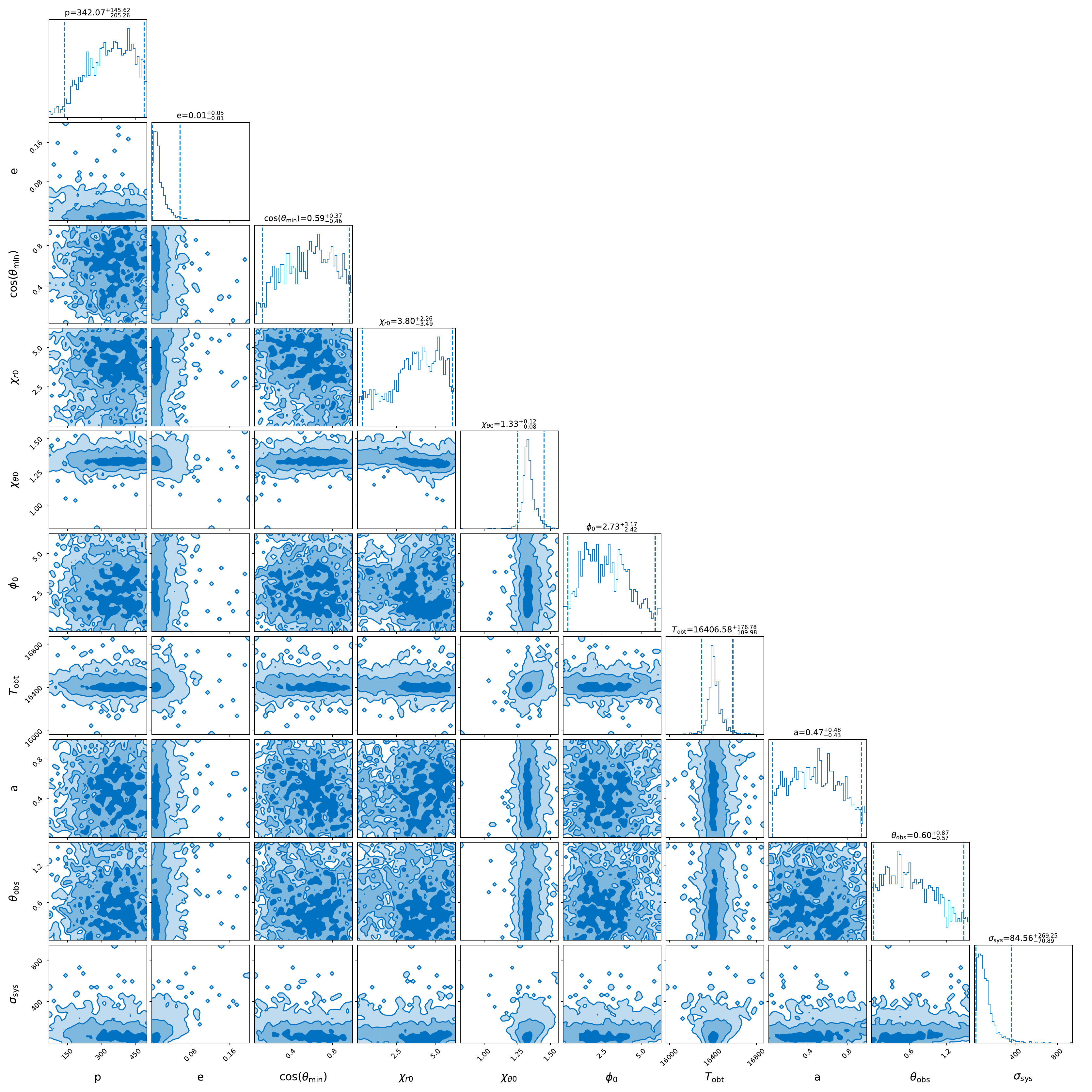}
\caption{\label{fig:eRO2_2_corner} Same to Fig.~\ref{fig:GSN_069_corner} except for eRO-QPE 2 (XMM2 and XMM3). }
\end{figure*}

\begin{figure*}
\includegraphics[scale=0.33]{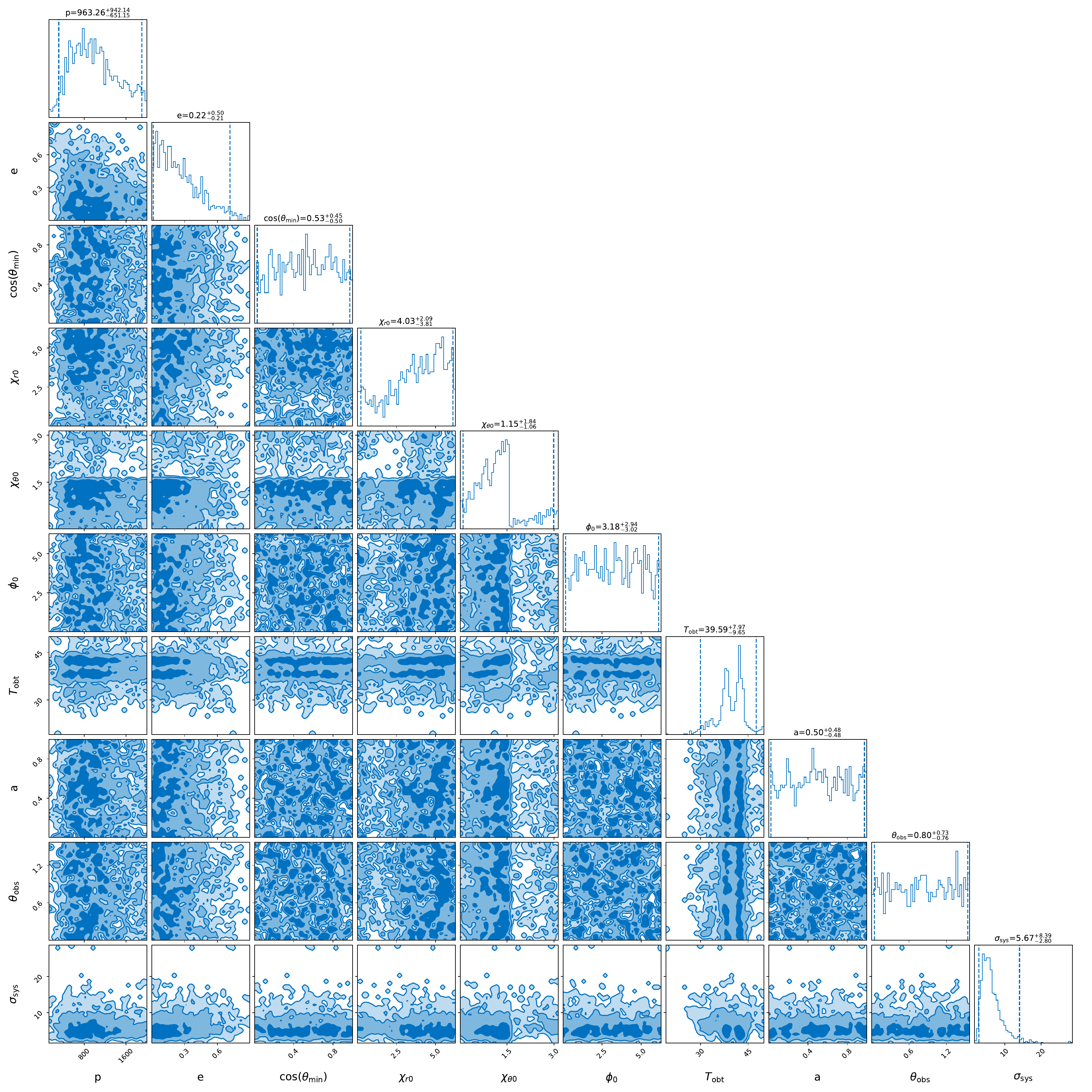}
\caption{\label{fig:Swift_corner} Same to Fig.~\ref{fig:GSN_069_corner} except for Swift J023017. }
\end{figure*}

\bibliography{ms}
\end{document}